\documentclass[preprint]{aastex}
 
\usepackage{amssymb}
\usepackage{epsfig}
\usepackage{amsmath}
 
\begin{document}

\title{Discovery of Reflection Nebulosity Around Five Vega-like Stars}
\email{}

 
\author{Paul Kalas\altaffilmark{1,2}, James R. Graham\altaffilmark{1,2}, Steven Beckwith\altaffilmark{3},}
\author{David C. Jewitt\altaffilmark{4} and James P. Lloyd\altaffilmark{1,2}}
\email{kalas@astron.berkeley.edu}
 
 
\altaffiltext{1}{Astronomy Department, University of California,
601 Campbell Hall, Berkeley, CA 94720
}
\altaffiltext{2}{Center for Adaptive Optics,
University of California, Santa Cruz, CA, 95064
}
\altaffiltext{3}{Space Telescope Science Institute, 3700 San Martin Drive,
Baltimore, MD 21218
}
\altaffiltext{4}{University of Hawaii, 2680 Woodlawn Drive, Honolulu, HI 96822
}

\begin{abstract}
Coronagraphic optical observations of six Vega-like stars reveal 
reflection nebulosities, five of which were previously unknown.
The
nebulosities illuminated by HD 4881, HD 23362, HD 23680,
HD 26676, and HD 49662 resemble
that of the Pleiades, indicating an interstellar origin for dust
grains.
The reflection nebulosity
around HD 123160 has a double-arm morphology, but no disk-like
feature is seen as close as 2.5$\arcsec$ from the star
in K-band adaptive optics data.  We demonstrate that uniform density dust clouds
surrounding HD 23362, HD 23680 and HD 123160 can account for
the observed 12$-$100 $\mu$m spectral energy distributions.
For HD 4881, HD 26676, and HD 49662 an additional emission
source, such as from a circumstellar disk or non-equilibrium 
grain heating, is required to fit the 12$-$25 $\mu$m data.  
These results indicate that in some cases, particularly
for Vega-like stars located beyond the Local Bubble ($>$100 pc),
the dust responsible for excess thermal emission
may originate from the interstellar medium rather than from
a planetary debris system.

\end{abstract}
 
 

\section{Introduction}

The far-infrared (FIR), all-sky survey conducted by the Infrared
Astronomical Satellite (IRAS)
revealed that roughly 15\% of nearby main sequence
stars have excess thermal emission due to the presence of
circumstellar dust \citep{bag87, aum88, ple99}.  
The thermal excess was first discovered around
Vega \citep{aum84}, and approximately 100 main sequence stars display
the ``Vega Phenomenon'' \citep{bap93}.  
The spectral energy distributions (SEDs) of Vega-like stars
are typically fitted assuming disk-like distributions of
grains \citep{sas96}, though in the optically thin regime
the adopted geometry plays no role.
Assuming the grains orbit the host stars as circumstellar disks, 
the dust destruction timescales are
typically one or two orders of magnitude shorter than the stellar ages. 
The existence of unseen parent bodies, exosolar analogs to Kuiper Belt objects,
is inferred in order to
replenish the dust complex via collisional erosion \citep{bap93}.

High resolution imaging of Vega-like stars
has confirmed the existence of circumstellar debris disks around $\sim$7 main
sequence stars \citep{kal98, lag00}.  However, the ratio of debris disks 
inferred from the FIR SEDs, to
those spatially resolved by imaging, is roughly 10:1.  
Here we present coronagraphic images of reflection nebulosity
surrounding six main sequence stars previously identified
as Vega-like \citep{bap93,syl96}.  
The new data show dust
morphology that is not disk-like, but qualitatively resembles
the Pleiades reflection nebulosity in five cases.  Dust scattered light
detected around HD 123160 shares characteristics with nebulosities seen in
both the Pleiades
and in star forming regions.

\section{Observations \& Data Reduction}

We used an optical coronagraph at the University of Hawaii (UH) 2.2 m telescope
to obtain  CCD images in the $R$-band (Table 1).  HD 123160 was also
imaged in the $V$-band.  
Reimaging optics gave 0.4$\arcsec$ per pixel, sufficient to Nyquist
sample the $\sim$1.2$\arcsec$ full-width at half-maximum (FWHM)
point-spread function (PSF).  
The field of view was a circular region with diameter 5.5$\arcmin$. 

Follow-up observations were obtained in the near-infrared (NIR)
for HD 123160
using  the Lick Observatory 3 m telescope (Table 1).
A 256$\times$256 pixel (0.076$\arcsec$/pixel), 
coronagraphic, near-infrared camera was used to artificially 
eclipse the central star with a 0.7$\arcsec$ wide finger \citep{jpl00}.  
Adaptive optics compensation \citep{max97}
using the primary star as a wavefront reference source improved image
quality from $\sim$1$\arcsec$ intrinsic seeing to 
$\sim$0.16$\arcsec$ (FWHM).

Our target sample was compiled from lists of Vega-like stars 
published by \citet{syl96} and \citet{bap93}.
We imaged all 22 Vega-like stars from Table 16 in \citet{syl96}, 
and 38 of the 60 stars from Table VIII in
\citet{bap93}.  The total sample of 79 stars (the two tables have three stars
in common) includes several with resolved debris
disks ($\beta$ Pic, Vega, $\epsilon$ Eri, and Fomalhaut), two
members of the Pleiades (18 Tau and 21 Tau), and three
Herbig Ae/Be stars (HD 34282, HD 35187, and HD 141569).  
We also observed stars nearby in the sky and with
magnitudes comparable to the science targets to be used later as template stars
for PSF subtraction.

Data reduction followed the standard steps of bias subtraction, flat-fielding,
sky subtraction, image registration, and median-filtering of multiple frames to
attain the effective integration times listed in Table 1.  The PSF of each
science target was subtracted by using either template PSFs discussed above,
or by radially sampling the template PSF, fitting the data with a 7th order
polynomial, and producing an artificial PSF from this fit.  The PSF subtraction
step is useful for extracting faint nebulosity as close to the star as possible, but
the nebulosities discussed here are detected in raw data. Residual noise after
the PSF subtraction dominates the data $\sim$1$\arcsec$ beyond the edge of the occulting
spot.  Table 1 lists the radius centered on each target star below which we cannot
obtain reliable information about the circumstellar environment.

\section{Results}

The suppression of direct stellar light with the coronagraph reveals the presence
of faint reflection nebulosity around the six Vega-like stars in Table 1 and Fig. 1.
The nebulosites have the following general properties:  
(a) spatial extent $\sim$1$\arcmin$$-$2$\arcmin$ radius; 
(b) surface brightness varying as radius, $r^{-\beta}$, with $\beta$$\leq$ 2;   
(c) range of surface brightness approximately 
20 mag arcsec$^{-2}$ to 24 mag arcsec$^{-2}$; and
(d)  linear, filamentary, striated morphological structure, 
similar to that observed in the Pleiades reflection
nebulosities \citep{arn77}.
The background noise, 3$\sigma\sim$24 mag arcsec$^{-2}$, limits the radial 
extent to which
each nebulosity is detected.
Table 1 gives the $Hipparcos$ parallaxes, indicating heliocentric
distances $>$100 pc for every star.  The detected nebulosities therefore span
spatial scales 10$^3$ $-$ 10$^5$ AU from their parent stars, and the
occulted regions obscure the central $\sim$10$^3$ AU (Table 1).

The nebulosity around HD 26676 (Fig. 1D), the brightest and most extended,
has been previously documented by \citet{vdb66}
and identified as a member of the Tau R2 association \citep{rac68}, which is
part of the Pleiades.  An examination of the literature indicates that the
other five nebulosities were not previously known.  

The nebulosity around HD 123160 is characterized by  
a double-arm structure southeast of the 
star (Fig. 1F; Features $i$ and $ii$), rather than the linear
features seen around the other five stars.  Arm-like morphologies appear near
several pre-main sequence stars such as
GM Aur and Z CMa \citep{nak95}.
The outside boundary of arm $i$ traces a 
closed curve that intersects feature $iii$. The latter is
curved in the same direction as
$i$, indicating that $iii$
may be a detached segment of $i$.
Another diffuse, curved structure, $iv$, is 95$\arcsec$
from the star with position angle and curved morphology
comparable to feature $iii$.  Neither 
$iii$ nor $iv$ have a core+halo structure that characterizes
background field galaxies in these data.
We also detect two point sources lying along a north-south
axis and separated by
3.6$\arcsec$ at position $v$.
They are surrounded by a nebulosity that has a tail
pointing towards HD 123160.  Both are red, with
$V-K$=1.7 mag and 1.1 mag for the north and south
stars, respectively (after applying the extinction correction
determined for HD 123160 in Table 3).  The colors of the
northern component are consistent with a late-type stellar
photosphere.  However, the southern component
has $H-K$=1.1 mag, indicating an additional source of near-infrared
emission such as from a circumstellar disk \citep{meyer97}.

Figure 2 shows that
two brightness knots exist within arm $ii$. In the K-band adaptive optics
data, the north knot, $iia$, contains a point-like source 9.7 $\pm$0.2$\arcsec$ from the
star superimposed on the broader nebular emission.
After subtracting an estimate for the background contributed by
the broader nebulosity, the point-like source has $V$=16.5$\pm$0.3 mag and
$K$=14.2$\pm$0.3 mag (extinction corrected; Table 3). The $V-K$ color is
consistent with a K dwarf spectral type and distance modulus $>$100 pc.
Compared to the optical data, the adaptive optics $K$-band data probe
closer to the star for any structure that may have a disk-like morphology.  
No circumstellar
disk is detected as close as $\sim$2.5$\arcsec$ from the primary.  

Aperture photometry (diameter = 4.0$\arcsec$) performed on
the nebulosities labeled in Figs. 1F and 2
gives $R$ and $V$-band 
fluxes that decrease radially from the star with $\beta\leq$1, rather than
$\beta=$2 or $\beta=$3 that would characterize a homogenous
medium or dust disk, respectively.  For example, $iii$ has 9\% the flux
of $iia$, but is 5 times farther from the primary.  If both features
are produced by grains with similar scattering properties, then  
$iii$ has approximately twice as many
scattering grains as $iia$.  The $V-R$ colors range from 0.0 to -0.2 mag
after an extinction correction is applied.  Using $R$=7.81 mag
for HD 123160 \citep{syl96}, $V-R$ = +0.2 mag if HD 123160 is a
G5V star, or  $V-R$ = +0.3 mag if it is a G0III star.  Therefore the 
nebulous features in Figs. 1F and 2 appear bluer than HD 123160,
consistent with the scattered-light colors of the Merope nebula \citep{her96, her01}.

The discovery of nebulosity around these Vega-like stars that is not disk-like raises
the question of whether or not the dust emitting in the FIR is 
contained in disks related to planet building, or merely due to dust
contained in the ISM. 
The ``Pleiades Phenomenon'' is due to the chance encounter of bright, nearby
stars with a clump of ISM \citep{arn77, whi93, her01}.  
Filamentary features in optical scattered light are produced by the shearing of the cloud
as radiation pressure pushes dust around each star.
Our data do
not necessarily exclude the existence of circumstellar disks in addition
to the Pleiades-like nebulosity because the observations
are not sensitive to the central $\sim$4$\arcsec$ radius (Table 1).  
However, the main reason to presume the existence of a circumstellar disk
is that the IRAS data give fluxes in excess of photospheric levels.  If the
Pleiades-like dust detected here is capable of producing the
FIR emission, then the Pleiades Phenomenon is a more plausible explanation
for the FIR excesses than the Vega Phenomenon.  Below we test
the validity of attributing the observed excess FIR emission to an interstellar dust
cloud encountering each of our target stars.

\section{Thermal Emission Model for the Pleiades Phenomenon}

To test if the IRAS far-infrared data are consistent with the Pleiades Phenomenon, 
we experiment with a model comprising a star
embedded in a uniform number density, optically thin, dust cloud.  
The geometry is spherical, except that radiation pressure from the star excavates an
axisymmetric paraboloidal
cavity within the cloud, as described by \citet{art97}.  In spherical coordinates,
(r, $\theta$, $\phi$), a volume element is given as
dV = dr $\times$ r sin$\theta$d$\phi$ $\times$ r d$\theta$.
The inner boundary of the cloud
cavity has radius, 

\begin{equation}
r = r_{in} \biggl(\frac{2}{1+cos\theta}\biggr)
\end{equation}

The smallest radius that grains can approach the star, $r_{in}$, depends on
the relative velocity and the force of radiation pressure (which scales with stellar
luminosity and grain properties, such as size).  For example, at a relative velocity
of 12.2 km s$^{-1}$, grains $<$0.2 $\mu$m in size will not approach
Vega closer than $\sim$2000 AU, whereas 1 $\mu$m sized grains will
reach as close as $\sim$200 AU \citep{art97}.  In our simulations $r_{in}$ is
a free parameter.

Grain temperatures are calculated in each volume element 
assuming grains receive blackbody radiation from the star,
and emit as blackbodies.  Grain absorption and emission
efficiences will depend on grain properties
and wavelength regimes.  We estimate the absorption and emission efficiencies
empirically by assuming that they have the same 
functional form as the extinction
function for interstellar grains
given by \citet{mat90} in his
Table 1.  A smoothed version of this extinction function
can be described by three power laws with form $\epsilon(\nu) = \epsilon_o\nu^{\alpha}$,
where $\alpha$=0 for $\lambda < $ 0.5 $\mu$m, 
$\alpha$= 1.68 for 0.5 $\mu$m $\leq \lambda < $ 5 $\mu$m.
and $\alpha$=1.06 for $\lambda \geq $ 5 $\mu$m.  Very hot stars
will put out their energy at short wavelengths where the
grain absorption efficiency is at a maximum, whereas cool stars
will not efficiently heat these interstellar grains.
Most of the energy re-radiated by grains will occur in the $\lambda \geq $ 5 $\mu$m
regime, and therefore will have emission efficiency proportional to $\nu^{1.06}$.

The dust cloud thermal emission is fitted to the observed IRAS fluxes by
adjusting the grain number density, $\rho$, the inner minimum radius
of the cavity, r$_{in}$, and the outer radius of the cloud, R$_{out}$. 
The photospheric contribution from optical wavelengths through
12 $\mu$m is fit by adjusting the stellar temperature, radius, and
heliocentric distance.  In the first iteration the stellar temperature
and radius are taken from \citet{cox99} based on the spectral type
for each star given in the literature.

Table 2 gives the observed optical, near-infrared, and far-infrared magnitudes
and fluxes.  Interstellar reddening and extinction were calculated using
both the B$-$V and V$-$K colors and is given in Table 3;
we used R$_V$=3.1 for the total-to-selective
extinction values given by \citet{mat90} in his Table 1.  
The color correction for the FIR data was obtained from
Table VI.C.6 in the {\it IRAS Explanatory Supplement Version 2}.  In Table 4 we list the 
extinction-corrected optical and NIR photometry, and the IRAS fluxes color-corrected
based on the temperatures listed in Table 1.  In several cases an alternate
spectral type is proposed and the extinction and photometry are recalculated.
We considered both the
IRAS PSC and the FSC data for fitting the model to the 12$-$100 $\mu$m fluxes.  
The uncertainties for these
data are given in the respective IRAS catalogs, with typical values $\sim$10\%.
The heliocentric distances used
in the first iteration are taken from the $Hipparcos$ parallaxes for each star (Table 1),
but beyond 100 pc the $Hipparcos$ distances have significant uncertainties.

Table 5 gives two examples of model fits to the SED of each star, 
and Figs. 3 - 10 display the model fits as discussed below.  Table 5 also
lists the maximum grain temperature, T$_o$, the fractional infrared luminosity of the cloud, 
L$_{d}$/L$_{\star}$ and the maximum optical depth, $\tau_{max}$. 
For HD 4881, HD 26676 and HD 49662, the model fits to the 60$-$100 $\mu$m emission gives
a deficit of 12$-$25 $\mu$m emission.
Our simple model assumptions do not 
account for at least four factors that could enhance the 12$-$25 $\mu$m emission.

First, density variations exist within the dust cloud, as shown in the optical images (Fig. 1).
The theoretical model of \citet{art97} also shows that a ``snowplow'' effect
occurs as the dust moves around the cavity maintained by radiation pressure.
A skin of higher density material will exist near the surface of the cavity.
If we assume that the cavity is a sphere for simplicity, and displace the
material that would have occupied the sphere to an annular
skin with thickness $\Delta$r = r$_s$ - r$_{in}$, then the mass density
in the skin is:

\begin{equation}
\frac{\rho_{skin}}{\rho_{cloud}} = 1 + \biggl(\frac{r_{in}^3}{r_s^3-r_{in}^3}\biggr)
\end{equation}

The larger the dust cavity produced by the star, the denser the 
skin layer for a fixed skin thickness.  For example, $\rho$ is a factor of two greater
in the skin compared to the ambient cloud for r$_{in}$=200 AU 
and r$_s$=250 AU, and a factor of 3 greater if we assume r$_{in}$=500 AU and r$_s$=550 AU.  
We find that including a high density surface layer in the model increases 
the 12$-$25 $\mu$m emission slightly, but
a good fit to the data requires a factor of $\sim$10$^2$ increase in dust density 
between the skin and the ambient material.  Thus, the enhanced 12$-$25 $\mu$m emission must
have additional sources. 

Second, larger grains ($\geq$1 $\mu$m) will not
be pushed back by radiation pressure and will occupy the cavity \citep{art97}.
We find that filling the cavity with grains also does not succeed in
fitting the observed 12$-$25 $\mu$m emission, given no
increase in the density of material above the ambient medium (Table 5).  

Third, the smallest grains ($\leq$0.1 $\mu$m) will undergo
non-thermal heating events to $\sim$1000 K \citep{gre68, sel84}.  The empirical finding for
the Pleiades is that
the 12$-$25 $\mu$m emission should represent $\sim$30\% of the total
nebular FIR flux \citep{cas87, sel90}.  The IRAS SED's for four stars in the Pleiades
show that the 12$-$25 $\mu$m flux densities
lie significantly above a blackbody fitting the 60$-$100 $\mu$m data \citep{cas87}.  
Thus, a hot grain component could be present
that will add 12$-$25 $\mu$m flux that is not accounted
for by our model.  However, the existence of small-grain heating
should also produce observable NIR excess emission and infrared emission features 
\citep{dbp90, sel96, ssb97}.

Fourth, our simulation does not take into account the existence of a circumstellar
disk in addition
to the ISM nebulosity detected in this paper.
Material close to the star with number density decreasing with
radius would add 12$-$25 $\mu$m flux to our model SED.
Future high resolution observations are required to detect
number density variations closer to the star than our observations
permit (Table 1). In four cases below we discuss previous attempts to
fit the SED's with model circumstellar disks.

\subsection{18 Tau and 21 Tau}

Before applying our model to the stars shown in Fig. 1, we test it on
two Pleiads that are in the IRAS PSC and FSC, and have been identified
as candidate Vega-like stars \citep{bap93}.  18 Tau (HD 23324) and 21 Tau (HD 23432) are both
B8V stars with 
$Hipparcos$ distances between 110 pc and 120 pc.  Two model fits to the SED of
each star are given in Figs. 3 and 4, and Table 5.  The parameters for models 18 Tau-a are  
chosen to produce a SED that
fits the PSC 100 $\mu$m data.  Model 18 Tau-b
demonstrates that a fit to the FSC 100 $\mu$m point requires decreasing the mass density, $\rho$, and increasing the outer
radius, $R_{out}$, relative to model 18 Tau-a.  However, no combination of parameters
can fit the 12$-$25 $\mu$m and the 60$-$100 $\mu$m
regions simultaneously, in agreement with the findings of \citet{cas87}.
These authors suggest that the non-equilibrium heating of small grains
may account for the observed excess flux in the IRAS 12 and 25 $\mu$m passbands.
The angular radii of models 18 Tau-a and 18 Tau-b would be 8$\arcmin$ and 21$\arcmin$, respectively.
\citet{gvb93} measure 8$\arcmin$ radius for 18 Tau in the IRAS Infrared Sky Survey Atlas (ISSA).  Thus,
model 18 Tau-a is preferred over 18 Tau-b.

For a straightforward comparison, the model fits to the 21 Tau data use the same stellar parameters
as the 18 Tau models.
Model 21 Tau-a fits the IRAS 60$-$100 $\mu$m PSC data and again 
demonstrates that a second source of 12$-$25 $\mu$m
emission is necessary.  Relative to the 18 Tau models, we decrease the inner radius to 10 AU to
show that the hotter grains do not add enough 12$-$25 $\mu$m flux to match the observations.
However, the 60 $\mu$m flux density for 21 Tau in the IRAS FSC is significantly higher 
than in the PSC.  Model 21 Tau-b gives a fit to the FSC data that agrees with the entire
observed 12-100 $\mu$m SED.  Thus, the errors in any single FIR data point may change the physical
interpretation significantly.  
The angular radii of models 21 Tau-a and 21 Tau-b would be 8$\arcmin$ and 3.7$\arcmin$, respectively.
\citet{gvb93} measure 5$\arcmin$ radius for 21 Tau in ISSA maps.  Thus,
model 21 Tau-b is preferred over 18 Tau-a.

\subsection{HD 4881}

Given the B9.5V spectral type for HD 4881, our model gives
optical-NIR flux densities that lie well
below the observed values if we assume the $Hipparcos$ distance of 350 pc (Table 1)
and T$_{eff}$=11,400 K.  Rather, the fit to the optical and NIR data shown in Fig. 5
assumes d=168 pc and T$_{eff}$=12,300 K.  The
higher temperature is consistent with the \citet{mir99} reclassification
of this star as B8.  The 25$-$100 $\mu$m fluxes 
are fitted with the dust cloud parameters given in Table 5.  Figure 5 maps the
spectral energy distributions of two model fits.  Fitting the 100 $\mu$m
photometry requires that R$_{out}$ is not smaller than $\sim$8.7$\times$10$^4$ AU.
At the assumed distance, the model dust cloud subtends $\sim$8.6$\arcmin$ radius,
which is approximately equal to the average angular extent measured by \citet{dri96}
in IRAS Skyflux plates.  This is larger than the $\sim$5$\arcmin$ radial
extent measured by \citet{gvb93} and \citet{mir99}, but a factor of two
uncertainty in the measured angular extent has been demonstrated by \citet{jur99}. 
We find that r$_{in}$ may be varied between
1 AU and 800 AU with the resulting 25 $\mu$m flux density contained within the
IRAS PSC and FSC data points, and with negligible effect on the
12 $\mu$m flux density.  

At 12 $\mu$m our
model photosphere+cloud gives F$_{12}$=0.13 Jy, whereas the observed
IRAS values are 0.19 $\pm$ 0.02 Jy in the PSC, and 0.25 $\pm$ 0.02 Jy in
the FSC.  A poor fit to the 12 $\mu$m flux
was previously found in circumstellar disk models for HD 4881 \citep{cou98}, as
well as models that assume spherical dust clouds
\citep{mir99}. Here the 0.06 Jy difference between the flux density in our model and
the PSC flux density is equal to the disagreement between the PSC and FSC flux densities.
The statistical significance of the poor model fit is therefore best 
evaluated after follow-up 10$-$20 $\mu$m observations can better constrain
the photometry.  If the IRAS fluxes are confirmed, then a warm-grain
component needs to be added to the model, such as from a circumstellar
disk or nonequilibrium small-grain heating.

\subsection{HD 23362}

Assuming a K2V spectral type, \citet{syl96} determine that the photometric
distance to HD 23362 is 6.5 pc.  This contrasts sharply against the
subsequent $Hippparcos$ parallax measurement that places HD 23362 at
352 pc.  The derived visual extinction of $\sim$2 mag (Table 3) is also
inconsistent with the \citet{syl96} distance.  In view of the
high reddening, \citet{syl96} comment
that the distance to HD 23362 should be determined independently
and that the star may be misclassified.  

In order to produce a closer match to the $Hipparcos$ distance, we re-calculate
the extinction values assuming the spectral type
is K2III, with T$_{eff}$=4200 and R=20R$_\sun$ (Tables 3 and 4).  With
this assumption we are then able to
fit the optical and NIR photometric data
with a stellar blackbody at 187 pc (Table 5, Fig. 6). 
The model dust cloud parameters are similar to those of HD 4881, except
that there is no disagreement between the model SED and the
12 $\mu$m IRAS flux (Fig. 6).

\citet{sas96} attempted to fit the SED with a model circumstellar disk, but
failed to fit the 100 $\mu$m flux by an order of magnitude.  
They concluded that the 100 $\mu$m is due to infrared cirrus.
Our model suggests that the 12 $\mu$m $-$ 25 $\mu$m emission is photospheric,
and that all of the 60 $\mu$m $-$ 100 $\mu$m
emission is attributable to interstellar dust.  Model 23362-b (Table 5, Fig. 6)
indicates that R$_{out}$ may be as small as 1.1$\times$10$^4$ AU and still 
fit the IRAS PSC 100 $\mu$m data.  In this case the cloud 
subtends $\sim$1$\arcmin$ radius and would be unresolved in the IRAS 60$-$100 $\mu$m data.  

\subsection{HD 23680}

As with HD 23362, the photometric distance for HD 23680 given the G5V
spectral type is $\sim$20 pc \citep{sas96}, but both the $Hipparcos$ parallax and the reddening
are consistent with d$\sim$200 pc.  We recalculate the reddening values
assuming HD 23680 has spectral type G5III, and using
T$_{eff}$=5050 and R=10R$_\sun$, we fit the stellar SED using
distance d=205 pc (Table 5, Fig. 7).  The cloud model (23680-a) gives satisfactory fits with
R$_{out}$ = 7$\times$10$^4$ AU, which subtends $\sim$6$\arcmin$ radius.  However,
the second model (23680-b), with R$_{out}$ = 4$\times$10$^4$ AU, also gives an SED
passing within the error bars of the
100 $\mu$m data point.  In this case, the cloud's $\sim$3$\arcmin$ radius may
be resolved in the 100 $\mu$m IRAS data.  In our optical data, resolved patches of nebulosity
are detected as far as 1.7$\arcmin$ from the star, particularly to the north-northeast, with 
surface brightness R = 24 mag arcsec$^{-2}$.  Inspection of the
IRAS Sky Survey Atlas at 100 $\mu$m shows an asymmetric morphology
extended $\sim$3$\arcmin$ north of the stellar position.
Thus, the smaller outer radius for
model 23680-b is preferred over model 23680-a.
As with HD 23362, \citet{sas96} could not fit the
IRAS data with a circumstellar disk model.  In particular, the circumstellar disk model could
not reproduce the 100 $\mu$m flux.  Rather, the simple cloud model demonstrated
here is consistent with the FIR data.

\subsection{HD 26676 \& HD 49662}

For both of these B stars the dust cloud model cannot fit the 12$-$25 $\mu$m
and 60$-$100 $\mu$m regions simultaneously, as was found for the Pleiads 18 Tau and
21 Tau.  In Table 5 and Figs. 8 and 9 we show 
two different models for fitting the 12$-$25 $\mu$m data separately from the 
60$-$100 $\mu$m data.  The models fitting the 12$-$25 $\mu$m emission require
inner boundaries extending closer to the stars, and with greater dust number densities,
than the dust cloud models that fit the 60$-$100 $\mu$m data.
A number
density distribution increasing toward the star, such as with a circumstellar
disk, could also enhance the
12 $\mu$m and 25 $\mu$m fluxes.  Thus it is possible that these stars have 
circumstellar disks and happen to be interacting with ISM.
Alternately, the excess 12$-$25 $\mu$m emission could originate from non-equilibrium
small-grain heating.  As discussed in Section 3, HD 26676 is physically associated
with the Pleiades.  \citet{cas87} demonstrated that nonequlibrium grain
heating in the Pleiades nebulosity
produces $\sim$30\% of the total 12$-$100 $\mu$m emission at 12 $\mu$m and
25 $\mu$m.  Models 26676-a and 49662-a in Figs. 8 and 9 generate only $\sim$10\%
of the total observed emission at 12$-$25 $\mu$m.  Thus, an added small-grain component
could alter the resulting SED to fit the data.
If this is true, then evidence
should also exist for NIR excess or infrared emission features \citep{sel96}.  We have
no NIR data for HD 26676, but the NIR data for HD 49662 is consistent
with a purely photospheric origin.  The scale of emission for 
models HD 26676-a and HD 49662-a, given the distances listed
in Table 5, is $\sim$5.5$\arcmin$, which is consistent with the $\sim$10$\arcmin$ radii
measured by \citet{gvb93} in IRAS Skyflux plates.

\subsection{HD 123160}

The photometric distance of 16 pc assuming that HD 123160 is a G5V star \citep{syl96}
is inconsistent
with the lack of parallax information from the $Hipparcos$ and $Gliese$
catalogs, and with the high extinction (Table 3). Though we confirm that the
optical and NIR data may be fit with a G5V star at 16 pc, a nearly equal model SED
is obtained by
assuming a G0III star at $\sim$110 pc. For either spectral type the visual
extinction exceeds 2 magnitudes (Table 3).  In Fig. 10 and Table 5 we present
dust cloud models assuming HD 123160 is a distant giant.
However, further study of this system is necessary to determine its evolutionary status.
Lithium abundance measurements suggest that 
HD 123160 is relatively young, with age $\sim$70 Myr \citep{dbr97}.  
This age is comparable to that of many Vega-like stars as well as members of the Pleiades.  
The nebulosities $iia$ and $iib$ (Fig. 2) may originate from the
same physical mechanisms that produce
the IC 349 nebulosity near 23 Tau in the Pleiades \citep{her96, her01}.
The semi-stellar appearance of $iia$ is similar to the main knot in
IC 349
\citep{her96, her01}, except that it is very red (Section 3).  
On the other hand, the entire complex of nebulous features
shown in Fig. 1F resembles a star-forming region where a young star illuminates
its natal dust cloud.  Feature $iia$ could be a K star associated with HD 123160.
The SED's produced by our Pleiades cloud model fit the 
observed 12$-$100 $\mu$m data points (Fig. 10), and the scale of
emission for the models is $\sim$1.5$\arcmin$ radius.  This corresponds
to the projected separation between HD 123160 and feature $iv$ (Fig. 4F).  
A circumstellar disk
model used by \citet{sas96}, assuming HD 123160 is a G5V star at 16 pc, also gives satisfactory
fits to the data. However, the present high resolution data (Fig. 2) show no
evidence for a circumstellar disk, and we therefore favor the Pleiades cloud model.

\section{Discussion}

\citet{bap93} cautioned that thermal emission from
reflection nebulosities such as in the Pleiades may 
appear similar to thermal emission
from Vega-like stars.  
Our simulations of interstellar grain emission demonstrate that the 
nebulae shown in Fig. 1 are capable of producing the excess
thermal emission observed by IRAS.  In three cases, a circumstellar disk or non-equilibrium 
small-grain heating may account for the 12$-$25 $\mu$m emission.  The latter mechanism is 
consistent with our current understanding of grain emission from
the Pleiades.

A general problem in interpreting SEDs is that the models typically have as many 
adjustable parameters as there are data points.  Good fits are not persuasive by 
themselves to determine the distribution of the dust, especially when the dust is 
optically thin to absorption and emission.  Unlike the SEDs of young stellar objects, 
the Vega-like stars have strongly peaked far-infrared flux densities consistent with 
dust with a small range of temperatures.  To interpret the dust distributions as 
lying in a single plane (disks) requires additional information, such as 
images of the scattered light.

Other authors have fitted the IRAS data for HD 4881, HD 23362, HD 23680,
and HD 123160 with circumstellar disk models,
and in some cases interpreted excess emission at 100 $\mu$m
as due to  infrared cirrus \citep{low84} in the background.  Given the
optical data and the results of our modeling, we argue that the infrared cirrus is local
to each star, appearing as the Pleiades Phenomenon in scattered light.
From the infrared standpoint, the term ``cirrus hotspot'' is used
to describe the local heating of ISM by a star. 
To qualify as cirrus hotspots, FIR emission must be extended on arcminute
scales with color temperatures between 25 K and 70 K \citep{gvb93}.
The early-type stars in the Pleiades, for example, appear as
cirrus hotspots in the IRAS data. 
The three B stars in our list, HD 4881, HD 26676,
and HD 49662, are also identified as
infrared cirrus hotspots \citep{gvb93, dri96}.  

Our three B stars, as well as 18 Tau and 21 Tau from the original sample,
are given as Vega excess stars by
\citet{bap93} in their Table VIII. 
We find that all but three of the remaining 29 B stars in the \citet{bap93} table
are also identified as cirrus hotspots by \citet{gvb93}.
The two different interpretations for the same FIR data illustrate
the difficulty in uniquely identifying the origin of dust that
produces excess thermal emission.  For instance, \citet{bap93}
identified Vega-like stars if the color temperature satisfied
30 K $< T <$ 500 K, which overlaps the color temperature criterion
adopted by \citet{gvb93}.
However,
the Vega-like stars with resolved debris disks and rings (e.g. $\beta$ Pic,
$\epsilon$ Eri, Fomalhaut, HR 4796A) have heliocentric distances
d$<$100 pc, whereas
the stars discussed here are located at d$>$100 pc.   
Thus the unique identification of thermal excess is problematic
for the more distant objects.  From our initial sample of 79
Vega-like stars, 72 have $Hipparcos$-detected distances, and
of these 43 (60\%) have d $>$ 100 pc.
In part, the source confusion is
a question of spatial resolution. Resolved observations from the optical to the
far-infrared are essential for determining the nature of circumstellar dust.
However, the Sun also lies within a relatively ISM-free bubble 65$-$250 pc in
radius \citep{sfe99}.  Thus, the Pleiades phenomenon naturally occurs with greater frequency
among the more distant stars.  

Though distance is a first order measure for the reliability of identifying the Vega Phenomenon,
the Local Bubble has a non-spherical geometry, giving specific
directions with respect to the Sun 
that are most likely to contain denser ISM that would produce the Pleiades
Phenomenon.  \citet{sfe99} use the NaI D-line doublet to measure absorption toward stars
within 300 pc of the Sun, and produce maps of neutral gas in three galactic projections.  
In Figs. 11, 12, and 13, we plot the locations of Vega-like stars using the same
same galactic projections as \citet{sfe99}.  We take the 60 Vega-like stars from
\citet{bap93} Table VIII (``Bright star catalog main-sequence stars with
Vega-like far-infrared excesses''), and 73 Vega-like stars
from \citet{man98} Table 2 (``Newly identified
candidate main-sequence stars with debris disks'').  Two stars (HD 73390 and HD 181296)
in \citet{man98}
are in the \citet{bap93} table, leaving a total sample of 131 stars.
From this sample
we select the 111 stars that have $Hipparcos$-detected distances.  Finally, a total
of 85 Vega-like stars lie within the three planes of reference defined
by \citet{sfe99}, and those within 300 pc of the Sun
are plotted in Figs. 11$-$13.  We overlay
two contours that trace the lowest and highest NaI D2 absorption mapped by \citet{sfe99}.  
The figure captions give more details.

Figures 11$-$13 show that the walls of the Local Bubble approach
the Sun $<$50 pc in certain directions, but are $>$100 pc distant in other
directions, particularly toward the North Galactic Pole (Fig. 13; named the
``Local Chimney'' by Welsh et al. 1999).  In the galactic plane view (Fig. 11)
the Local Bubble has maximum extend toward $l$=225$\degr$, which is also
the direction toward HD 49662.  However, HD 49662 is located
right on the wall of the high density gas.  Three more Vega-like stars in
Fig. 11 (HD 52140, HD 28149 and HD 108257) appear spatially associated
with local overdensities of gas.  

Figure 12 shows that HD 23680 is in the same general direction of HD 26676 and both
are within the region of high-density gas.  Also evident
is a group of five stars that trace the high-density wall at $l$=180$\degr$, $b\sim-30\degr$,
d$\sim$125 pc.  These Vega-like stars (HD 23324 = 18 Tau, HD 23432 = 21 Tau, 
HD 28149, HD 28375, HD 28978)
are associated with Taurus and the Pleiades.  Two more groups of five stars each appear in Fig. 12.  
At $l$=0$\degr$, $b\sim-30\degr$, d$\sim$50 pc we find HD 181296, HD 191089, 
HD 176638, HD 181864, and HD 181327.
These are not associated with gas, but HD 181296 and HD 181327 are members of the 
Tucanae Association
\citep{zuc00}.  Their youth and the lack of interstellar gas favors the Vega 
Phenomenon interpretation
of their FIR excesses.  A third group of Vega-like stars is evident 
at $l$=0$\degr$, $b\sim25\degr$, 100 pc $<$ d $<$ 150 pc
(HD 142096, HD 142165, HD 143018, HD 145263, and HD 145482).  These are at the distance and
direction of the Upper Scorpius subgroup of the Sco OB2 association \citep{zee99} which
is encompassed by a giant reflection nebula \citep{gor94}.  The association of
these stars with large-scale dust
and gas favors the Pleiades Phenomenon interpretation for the FIR
excesses.

Figure 13 shows that Vega-like stars may be detected at the greatest
distances with the least confusion from the ISM in the direction of 
the North Galactic Pole.  The figure also shows a group of four
stars at $l$=90$\degr$, $b\sim-30\degr$, d$\sim$20 pc (HD 39060 = $\beta$ Pic,
HD 41742, HD 53143, and HD 67199).  Their location in the Local
Bubble and their association with $\beta$ Pic favors the Vega
Phenomenon explanation for their FIR excesses.

Overall, the fraction of Vega-like stars that lie at or beyond the Local
Bubble wall (thin contour) is $>$50\%.  In Fig. 11 we plot 40 Vega-like stars, but only
five are in the Local Bubble.  The remainder are located at
or beyond the wall of low density gas, and 16 of these are in high density
gas regions.  Another 11 stars from the \citet{bap93}
and \citet{man98} tables would appear in Fig. 11 if we plotted distances
between 300 and 600 pc.  From the 30 Vega-like stars plotted in Fig. 12, 16 are in 
the high-density gas and 6 more lie between the high density and low density
walls.  In Fig. 13, 6 Vega-like stars are in the Local Bubble, 4 lie between
the low and high density walls, and 4 more are found in the high density regions.  

The post-$Hipparcos$ distance determinations are not only useful for 
judging the position of Vega-like stars relative to the Local Bubble, they
also help evaluate the 
physical scale of thermal emission.  We found that the $Hipparcos$ parallaxes
for HD 23362, HD 23680, and HD 123160, as well as the large reddening values,
probably place these stars beyond 100 pc, rather than 6$-$16 pc, as determined in
pre-$Hipparcos$ investigations.  
The greater distances therefore
explain how the thermal emission may originate from ISM material extending
10$^3$$-$10$^5$ AU from each star and still meet the criteria for
inclusion in the IRAS PSC.  
The possibility that these stars are distant giants 
producing the Pleiades Phenomenon may help explain the infrared
excesses observed around $\sim$100 luminosity class III stars \citep{zuck95, jur99, kim01}.

\section{Summary}

We detect optical reflection nebulosity around six main-sequence stars that are
candidates for having debris disks.  Five nebulae share the morphological characteristics of
dust surrounding bright stars in the Pleiades. 
The environment of HD 123160 has features that resemble 
both the Pleiades
and star forming regions.
No disk-like structures are detected, though
our optical coronagraphic technique does not probe the circumstellar environment closer than 
$\sim$4$\arcsec$ radius.
The sensitivity-limited radii of the nebulosities are between 1$\arcmin$ and 2$\arcmin$ and
the radial measurements of surface brightness are consistent with uniform density dust clouds
illuminated by the central star.  

We show that thermal emission from an optically thin, uniform density dust cloud surrounding
HD 23362, HD 23680, and HD 123160 can entirely explain the 12$-$100 $\mu$m emission detected by IRAS.
The Pleiades Phenomenon, a random encounter between a clump of ISM and a star,
is the most likely explanation for the excess FIR emission.  For
HD 4881, HD 26676, and HD 49662, the blackbody cloud model cannot simultaneously
fit the 12$-$25 $\mu$m and 60$-$100 $\mu$m regions of the IRAS-detected SED.
These stars may have circumstellar disks in addition to the interstellar dust detected 
in the optical.  However, the excess 12$-$25 $\mu$m emission may also arise
from non-equilibrium heating of small grains.  We show that 18 Tau and 21 Tau in
the Pleiades have comparable SED's with signatures of hot grains in
the 12$-$25 $\mu$m fluxes. 
Future observations sensitive to
disk-like structure within 500 AU of each star, and that search for NIR excesses
and emission features, are necessary to determine if these stars are manifesting
the Vega Phenomenon and the Pleiades Phenomenon simultaneously.  

We find that most Vega-like B stars have also been associated with FIR cirrus
hotspots.  We demonstrate
that $>$50\% of Vega Phenomenon stars are located beyond the gas-poor Local Bubble, and many
are spatially associated with regions of high density neutral gas.  Thus,
a significant fraction of Vega Phenomenon stars beyond 100 pc may be confused
with Pleiades Phenomenon stars.  

%
%
%
{\bf Acknowledgements:}  
We are grateful to J. Gradie, B. Zuckerman, and E. Becklin
for access to their coronagraph. This work was supported in part by NASA
grants to DCJ and PK, and by the NSF 
Center for Adaptive Optics, managed 
by UC Santa Cruz under cooperative 
agreement AST-9876783.  This research has made use of the NASA/IPAC Infrared Science Archive, 
operated by JPL, California Institute of Technology.

\clearpage


\begin{table}
\small
\begin{center} 
\caption{Summary of Observations}
\begin{tabular}[h]{ccccccccc}
\hline \hline\\
Name  	&   Date  &Telescope& $\lambda$ & $R_{in}^{*}$ & Integration & SpT & Parallax&Nebulosity\\
	&   (UT)  &  &($\mu$m) & (arcsec)     &  (sec)      &     &  (mas)  &\\
\hline
HD 4881		&10/12/93&UH 2.2 m& 0.65      &	4.0	& 640	       &B9.5V&  2.84 & Pleiades\\
HD 23362	&01/30/00&...& 0.65      &	4.5	& 285         &K2V  &  3.24 &Pleiades\\
HD 23680 	&01/30/00&...& 0.65      &	4.5	& 600	      &G5V   &  5.54 &Pleiades \\
HD 26676 	&10/12/93&...& 0.65      &	4.0	& 320         &B8Vn &  6.49 &Pleiades\\
HD 49662 	&01/30/00&...& 0.65      &	6.7	& 640         &B7IV &  5.37 &Pleiades\\
HD 123160	&01/29/00&...& 0.65      &	4.1	& 510         &G5V  &  -0.01$^{\dagger}$&Double Arm\\
...		&01/30/00&...& 0.55      &	3.7	& 120         & ...   & ...     & ...\\	
...		&06/17/00&Lick 3.0 m& 2.15    &	2.5	& 360	       & ...   & ...     & ...\\
...		&06/03/01& ...  & 2.15    &	4.0	& 720	       & ...  & ...     & ...\\
...		&06/03/01& ...  & 1.66   &	4.0	& 600	       & ...   & ...    & ...\\
\hline 
\end{tabular}
$^{*}$Radius from target star blocked by coronagraphic occulting spot and PSF residuals.\\
$^{\dagger}$Negative value in {\it Hipparcos Catalog} indicates parallax measurement that is
smaller than the error.  In section 4.6 we estimate d$\sim$110 pc.
\end{center}
\end{table}

\begin{table}
\small
\begin{center}
\caption{Optical through Far-Infrared Fluxes (not dereddened or color-corrected)*}
\begin{tabular}[h]{ccccccccccccc}
\hline \hline\\
$\lambda(\mu$m) =&0.44 &0.55 & 0.88& 1.22& 1.65& 2.18& 3.55& 4.77&  12  &  25  &  60  & 100   \\
                &(mag)&(mag)&(mag) &(mag)&(mag)&(mag)&(mag)&(mag)& (Jy) & (Jy) & (Jy) & (Jy)  \\
\hline
HD 4881 	&  -  &  -  &  -  & -   &  -  & -   &  -   &   -  & 0.28 & 0.28 & 3.88 & 11.2 \\
		& 6.25& 6.22& 6.17& 5.99&5.99 &5.96 &  -   &   -  & 0.36 & 0.39 & 4.75 & 11.5  \\
HD 23362	& 9.53& 7.85& 6.07&   - &  -  & -   & 3.72& 3.98  & 1.42 & 0.50 & 0.67 & 2.88 \\
		& 9.62& 7.91& 6.18& 4.89& 4.05&3.85 &  -   &   -  & 1.38 & 0.44 & 0.72 & 6.32  \\
HD 23680	& 9.40&	8.60& -   & 6.01& 5.37& 5.24& 5.16 & 5.29 & 0.44 &  -   & 1.89 & 6.27 \\
		& 9.62& 8.39& 7.20&  -  &   - &   - &  -   &   -  & 0.43 & 0.19 & 2.08 & 6.02 \\
HD 26676	&  -  &  -  &   - &  -  &  -  &  -  &  -   &  -   & 1.15 & 6.43 & 26.3 & 37.2 \\
		& 6.26& 6.24& 6.20&  -  &   - &   - &  -   &   -  & 1.49 & 7.41 & 30.0 & 47.4 \\
HD 49662	& 5.30& 5.40&  -  & 5.7 & 5.5 & 5.6 & 5.8  &  -   & 0.43 & 1.62 & 4.60 & 5.68 \\
		& 5.29& 5.39& 5.47&  -  & 5.61& 5.59&  -   &   -  & - & - & - & \\
HD 123160	&10.12& 8.62& 6.94& 5.86& 5.07& 4.87& 4.73 & 5.33 & 0.60 & 0.38 & 3.03 & 3.97 \\
		&10.25& 8.66& 7.05&  -  &   - &   - &  -   &   -  & 0.62 & 0.37 & 3.11 & 4.41  \\
18 Tau		&  -  & -   & -   &  -  &   - &   - &  -   &   -  & 0.47 & 0.59 & 2.99 & 6.01 \\
		& 5.60& 5.66& 5.69&  -  &   - &   - &  -   &   -  & 0.42 & 0.74 & 3.52 & 12.80\\
21 Tau		&  -  & -   & -   &  -  &   - &   - &  -   &   -  & 0.42 & 1.14 & 4.68 & 10.90 \\
		& 5.72& 5.76& 5.74&  -  &   - &   - &  -   &   -  & 0.39 & 1.06 & 13.20 & 9.75\\
\hline
\end{tabular}
*0.44 $-$ 0.88 $\mu$m from \citet{syl96} (first row) and $Hipparcos$ (second row), 
1.22 $-$ 2.18 $\mu$m from  \citet{syl96} (first row) and 2MASS (second row), 
3.55 $-$ 4.77 $\mu$m from \citet{syl96},
12 $-$ 100 $\mu$m fluxes from IRAS PSC (first row) and IRAS FSC (second row)
\end{center}
\end{table}

\begin{table}
\small
\begin{center}
\caption{Extinction assuming A$_V$=3.0E($B-V$) and A$_V$=1.1E($V-K$)*}
\begin{tabular}[h]{cccccccccc}
\hline \hline\\
Name     & SpT & $B-V$&($B-V$)$_o$&E($B-V$)&$V-K$&($V-K$)$_o$&E($V-K$)&A$_V$ &A$_V$ \\
         &     &Obs.&         &      &Obs.&         &      &($B-V$) &($V-K$) \\
\hline
HD 4881	&B9.5V& 0.03 & -0.07 & 0.10 & 0.26 & -0.13 & 0.39 & 0.30 & 0.43 \\
	& B8V &  ...   & -0.11 & 0.14 & ...    & -0.24 & 0.50 & 0.42 & 0.55 \\
HD 23362 & K2V & 1.70 &  0.91 & 0.79 & 4.03 &  2.22 & 1.81 & 2.37 & 1.99 \\
	&K2III&  ...   &  1.16 & 0.54 &  ...   &  2.70 & 1.33 & 1.62 & 1.46 \\
HD 23680 & G5V & 1.01 &  0.68 & 0.33 & 3.26 &  1.59 & 1.67 & 0.99 & 1.84 \\
	&G5III&  ...   &  0.86 & 0.15 & ... &  2.10 & 1.16 & 0.45 & 1.28 \\
HD 26676 &B8Vn & 0.02 & -0.11 & 0.13 & -    &  -    &  -   & 0.39 &   -  \\
HD 49662 &B7IV &-0.10 & -0.14 & 0.04 &-0.20 & -0.29 & 0.09 & 0.12 & 0.10 \\
HD 123160&G5V  & 1.55 & 0.68  & 0.87 & 3.77 &  1.59 & 2.18 & 2.61 & 2.40 \\
	&G0III& ...    &  -    &  -   &  ...   &  1.75 & 2.02 &   -  & 2.22 \\
18 Tau	& B8V & -0.06& -0.11 & 0.05 &  -   &    -  &  -   &  0.15&  -   \\
21 Tau 	& B8V & -0.04& -0.11 & 0.07 &  -   &    -  &  -   &  0.21&  -   \\
\hline
\end{tabular}
{\center *The observed magnitudes used here are the average of the two values
given for each passband in Table 2.  A$_V$ from $V-K$ was used to determine A$_\lambda$ for
Table 4 except
when $V-K$ data not available.} 
\end{center}
\end{table}

\begin{table}
\small
\begin{center}
\caption{De-reddened optical and NIR magnitudes and color-corrected FIR fluxes*}
\begin{tabular}[h]{rcccccccccccc}
\hline \hline\\
$\lambda(\mu$m)=&0.44 &0.55 & 0.88& 1.22& 1.65& 2.18& 3.55& 4.77&  12  &  25  &  60  & 100  \\
                &(mag)&(mag)&(mag)&(mag)&(mag)&(mag)&(mag)&(mag)& (Jy) & (Jy) & (Jy) & (Jy) \\
\hline
HD 4881 (B9.5V) &  -  &  -  &  -  & -   &  -  &  -   & -   &   -  & 0.19& 0.20& 2.94 & 10.28 \\
		&5.69 & 5.79& 5.97&5.87 &5.92 & 5.91 & -   &   -  & 0.25& 0.27& 3.60 & 10.55 \\
(B8V)&  -  & -   & -   &  -  &  -  &  -   & -   &  -   &   - &   - &  -   &  -    \\
		&5.54 & 5.67& 5.91&5.84 &5.90 & 5.90 & -   &  -   &   - &   - &  -   &  -    \\
HD 23362 (K2V)  &6.90 & 5.86& 5.12& -   & -   & -    &3.63 & 3.93 & 1.00& 0.36& 0.51 & 2.64  \\
		&6.99 & 6.45& 5.23&4.33 &3.70 & 3.64 &  -  &  -   & 0.97& 0.31& 0.54 & 5.80  \\
\phm{xxx}(K2III)&7.60 & 6.39& 5.37& -   &  -  &  -   &3.65 & 3.94 &  -  &  -  &  -   &  -    \\
		&7.69 & 6.45& 5.48&4.48 &3.79 & 3.69 &  -  &  -   &  -  &  -  &  -   &  -   \\
HD 23680 (G5V)  &6.96 & 6.76&  -  &5.49 &5.05 & 5.04 &5.07 & 5.24 & 0.31&  -  & 1.43 & 5.75 \\
		&7.18 & 6.55& 6.32&  -  &  -  &  -   & -   &  -   & 0.30& 0.13& 1.58 & 5.52 \\
\phm{xxx}(G5III)&7.71 & 7.32& -   & 5.65& 5.15& 5.10 & 5.10& 5.26 &  -  &  -  &  -   &   -  \\
		&7.93 & 7.11& 6.59&  -  &  -  &  -   &  -  &  -   &  -  &  -  &  -   &  -   \\
HD 26676 (B8Vn) &  -  &   - &  -  &  -  &  -  &  -   &  -  &   -  & 0.79& 4.56& 19.92& 34.13  \\
		&5.74 & 5.85& 6.01&  -  &  -  &  -   &  -  &  -   & 1.03& 5.26& 22.73& 43.49  \\
HD 49662 (B7IV) &5.16 & 5.30&   - &5.67 & 5.48& 5.59 & 5.79&   -  & 0.30& 1.15& 3.48 & 5.21 \\
		&5.15 & 5.29& 5.42&  -  & 5.59& 5.58 &  -  &  -   &  -  &  -  &  -   &  -  \\
HD 123160 (G5V) &6.92 & 6.22& 5.78&5.18 &4.68 & 4.61& 4.61 & 5.27 &0.42 & 0.27& 2.30 & 3.64 \\
		&7.05 & 6.26& 5.89& -   & -   &  -  &  -   &  -   &0.43 & 0.26& 2.36 & 4.04 \\
\phm{xxx}(G5III)&7.16 & 6.40& 5.87& 5.23& 4.68& 4.63& 4.62 & 5.27 &  -  &  -  &  -   &   -  \\
		&7.29 & 6.44& 5.98&  -  &  -  &  -  &  -   &  -   &   - &   - &  -   &  -    \\
18 Tau (B8V)	&  -  & -   & -   &  -  &   - &   - &  -   &   -  & 0.33 & 0.42 & 2.27 & 5.51 \\
		& 5.40& 5.51& 5.62&  -  &   - &   - &  -   &   -  & 0.29 & 0.53 & 2.67 & 11.74\\
21 Tau (B8V)	&  -  & -   & -   &  -  &   - &   - &  -   &   -  & 0.29 & 0.81 & 3.55 & 10.00 \\
		& 5.62& 5.55& 5.64&  -  &   - &   - &  -   &   -  & 0.27 & 0.75 & 10.00 & 8.94\\
\hline
\end{tabular}
{\center *Table rows have same format as Table 2, except that a second spectral type is
given for several stars.  The change in the FIR color correction for the second spectral
type is negligible and not listed.}
\end{center}
\end{table}

\begin{table}
\small
\begin{center}
\caption{Pleiades Model:  Uniform density dust cloud}
\begin{tabular}[h]{rccccccccc}
\hline \hline\\
Star - Model&T$_{eff}$&d&R$_\star$ ($\times$10$^{11}$)&$\rho$ ($\times$10$^{-23}$)&r$_{in}$&R$_{out}$&T$_o$&$\tau_{max}$&L$_d$/L$_\star$\\
                &    (K)  &(pc)&    (cm)     &(g cm$^{-3}$)                  &    (AU)  & (AU)  & (K) & &       \\
\hline
HD 4881-a       &12,300 & 168   & 2.09  & 0.27& 800  & 10$^5$ 		& 146 & 1.6$\times$10$^{-3}$ & 1.9$\times$10$^{-3}$ \\
\phm{HD 4881}-b &  ...	& ...	& ...	& 0.29& 300  & 8.7$\times$10$^4$& 217 & 1.5$\times$10$^{-3}$ & 1.8$\times$10$^{-3}$ \\
HD 23362-a	& 4200	& 187	& 13.92 & 0.37& 400  & 2.0$\times$10$^5$& 148 & 4.5$\times$10$^{-3}$ & 2.0$\times$10$^{-3}$ \\
\phm{HD 23362}-b&  ...  & ...	& ...	& 0.38& 40   & 1.1$\times$10$^4$& 366 & 2.6$\times$10$^{-3}$ & 1.2$\times$10$^{-3}$ \\
HD 23680-a	& 5050	& 205 	& 6.97 	& 2.30&	200  & 7.0$\times$10$^4$& 176 & 9.8$\times$10$^{-3}$ & 4.7$\times$10$^{-3}$ \\
\phm{HD 23680}-b&  ...	& ...	& ...	& 3.20& 400  & 4.0$\times$10$^4$& 134 & 7.7$\times$10$^{-3}$ & 4.3$\times$10$^{-3}$ \\
HD 26676-a	& 11,600& 163	& 2.09 	& 2.50& 100  & 5.5$\times$10$^4$& 320 & 8.4$\times$10$^{-3}$ & 1.0$\times$10$^{-2}$ \\ 
\phm{HD 26676}-b&  ...	& ...	& ...	& 22.0&  50  & 3.0$\times$10$^3$& 421 & 4.0$\times$10$^{-3}$ & 4.4$\times$10$^{-2}$ \\ 
HD 49662-a	& 14,000& 180	& 2.47  & 0.13& 100  & 6.2$\times$10$^4$& 397 & 4.9$\times$10$^{-4}$ & 5.9$\times$10$^{-4}$ \\
\phm{HD 49662}-b&  ...	& ...	& ...	& 0.70&  10  & 1.1$\times$10$^4$& 1000& 4.8$\times$10$^{-4}$ & 5.7$\times$10$^{-4}$ \\
HD 123160-a	& 5,500 & 110	& 4.18	& 4.9 & 300  & 1.2$\times$10$^4$& 135 & 3.5$\times$10$^{-3}$ & 2.1$\times$10$^{-3}$ \\
\phm{HD 123160}-b&  ...	& 120	& ...	& ... & 200  & 1.1$\times$10$^4$& 158 & 3.2$\times$10$^{-3}$ & 2.0$\times$10$^{-3}$ \\
18 Tau-a	& 11,400& 125 	& 2.00 	& 0.21&	200  & 6.0$\times$10$^4$& 236 & 7.7$\times$10$^{-4}$ & 2.2$\times$10$^{-5}$ \\
\phm{18 Tau}-b&  ...	& ...	& ...	& 0.18& 200  & 1.6$\times$10$^5$& 236 & 1.8$\times$10$^{-3}$ & 1.9$\times$10$^{-5}$ \\
21 Tau-a	& 11,400& 125 	& 2.00 	& 0.32&	10   & 6.0$\times$10$^4$& 770 & 1.2$\times$10$^{-3}$ & 4.3$\times$10$^{-5}$ \\
\phm{21 Tau}-b&  ...	& ...	& ...	& 1.0 & 10   & 2.8$\times$10$^4$& 770 & 1.7$\times$10$^{-3}$ & 1.3$\times$10$^{-4}$ \\
\hline
\end{tabular}
\end{center}
\end{table}

\newpage
\clearpage
\begin{figure}
\begin{center}
\psfig{file=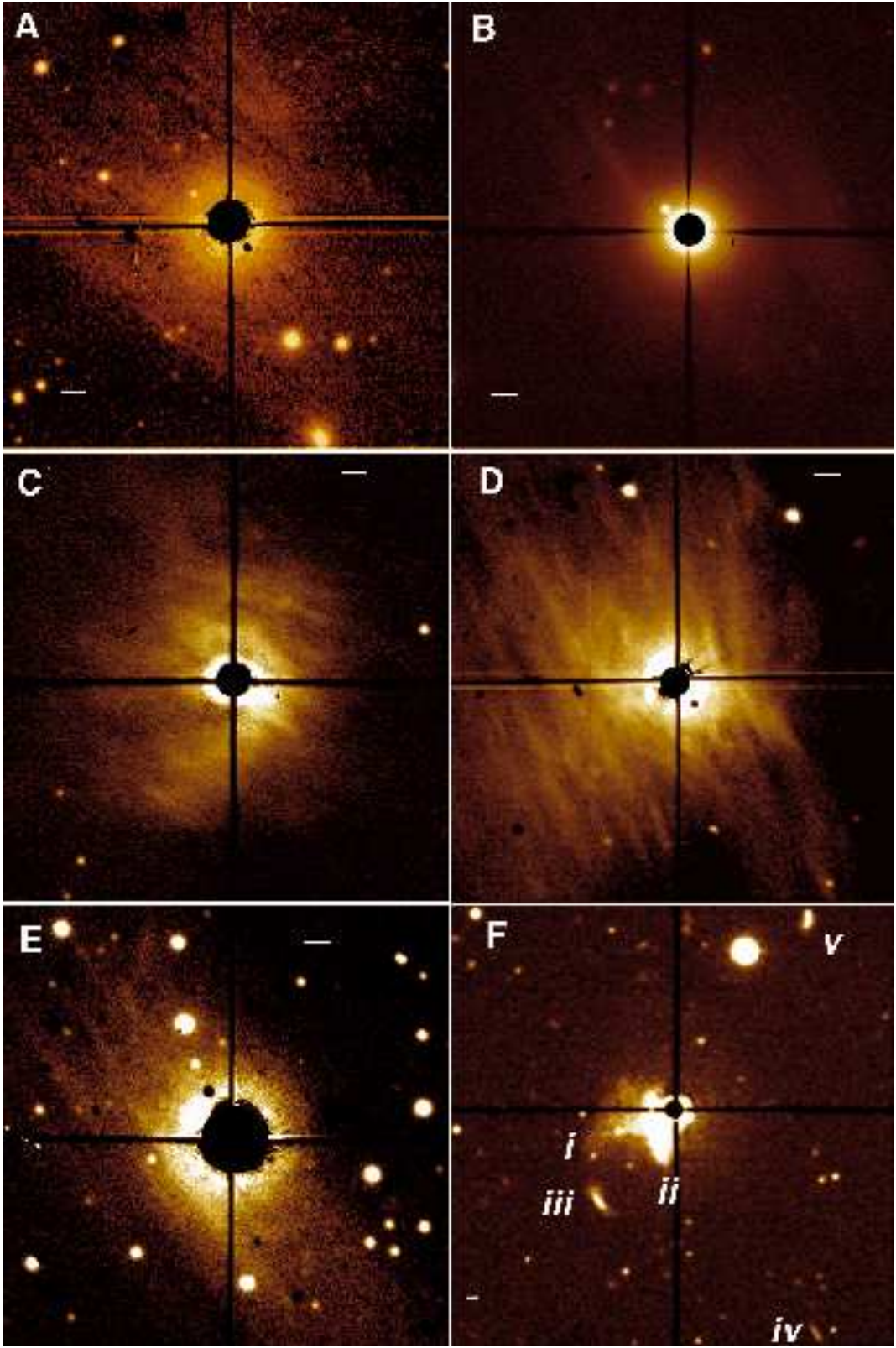}
\caption{R-band coronagraphic images of six Vega-like
stars after PSF subtraction [A: HD 4881, B: HD 23362
C: HD 23680, D: HD 26676, E: HD 49662, F: HD 123160].
North is up, east is left,  
the white bar represents 5$\arcsec$, and each box is 
1.5$\arcmin$ on a side, except for F which is
3.0$\arcmin$ on a side.   
For F we label five regions of nebulosity
discussed in the text.  Negative features are
artifacts of the PSF subtraction step.}
\end{center}
\end{figure}

\clearpage
\begin{figure}
\plotone{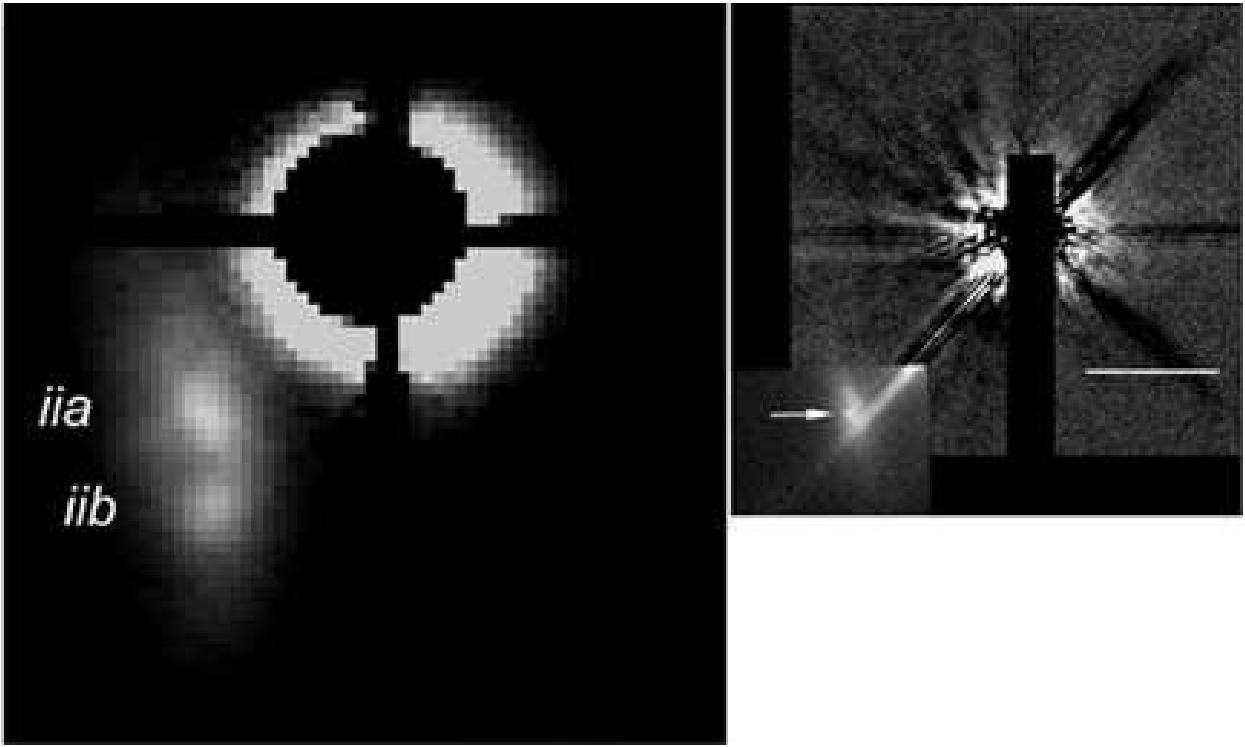}
\caption{The left panel is a different grayscale stretch 
and field of view of
Fig. 1F, showing the existence of 2 brightness peaks 
within arm $ii$.  The right panel shows the same
field imaged with adaptive optics at 2.15 $\mu$m on
two different nights.  The wide field shows the 
17 June 2000 data where HD 123160 was placed behind
the occulting finger and a PSF template was subtracted
to search for disk-like nebulosity.  In the lower left of this frame
we superimpose the 03 June 2001 $K$-band data which has
higher signal-to-noise due to the greater integration
time (Table 1) and the fact that no PSF template is
subtracted.  In both $K$-band data sets, as well as the $H$-band
data, feature $iia$ is detected as a possible point source at PA$\sim$135$\degr$,
indicated here with an arrow.  
The white bar represents 5$\arcsec$.
\label{fig2}}
\end{figure}

\clearpage
\begin{figure}
\plotone{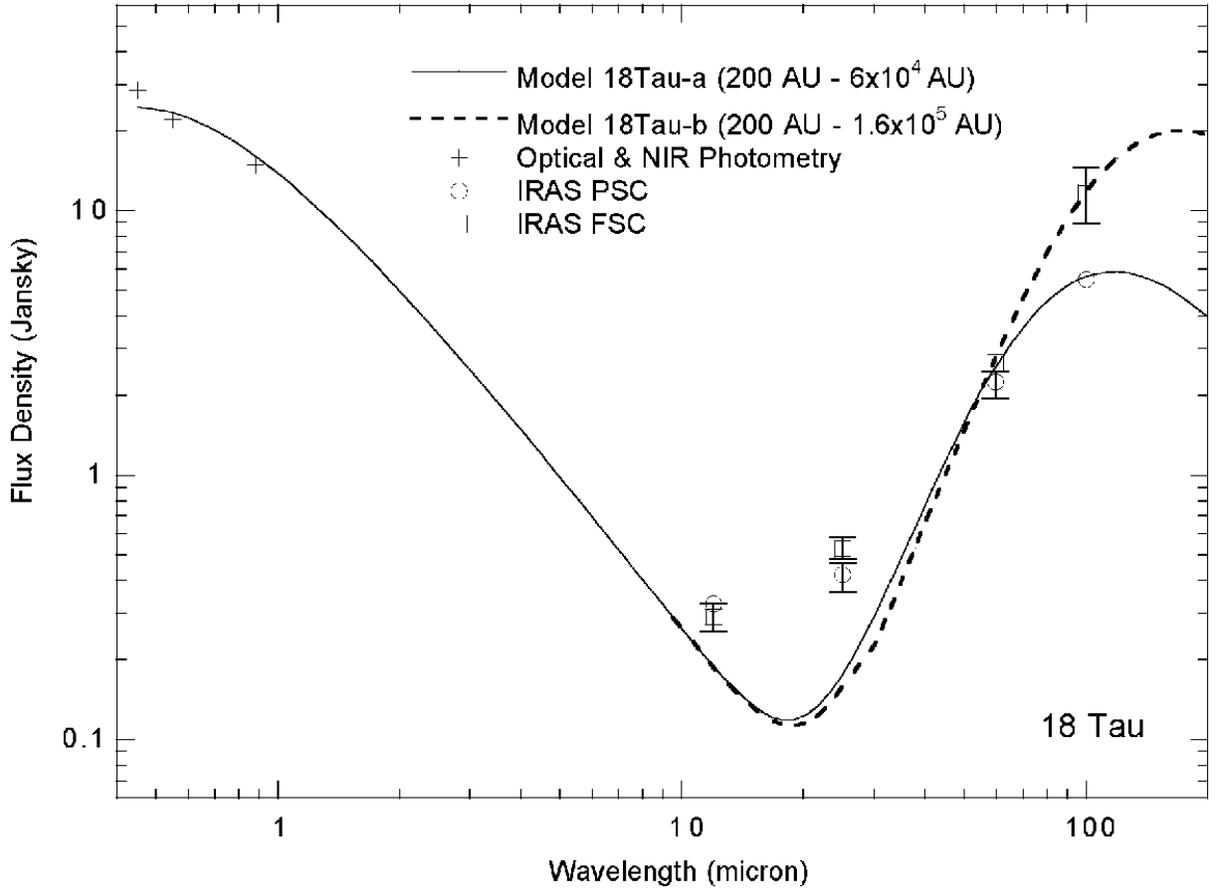}
\caption{Photometric data for 18 Tau (Table 4) and spectral energy distributions from
dust cloud models (Table 5).  Symbol sizes for the IRAS data approximate the sizes of error bars
unless marked otherwise. 
The two models demonstrate different cloud properties that give fits to either the PSC or
FSC 100 $\mu$m data.  However, the model cannot fit the 12-25 $\mu$m and 60-100 $\mu$m
regions simultaneously, confirming previous findings \citep{cas87}.
Non-equilibrium heating of small grains is thought to produce flux in
excess of the 12-25 $\mu$m blackbody emission simulated here.
\label{fig3}}
\end{figure}

\clearpage
\begin{figure}
\plotone{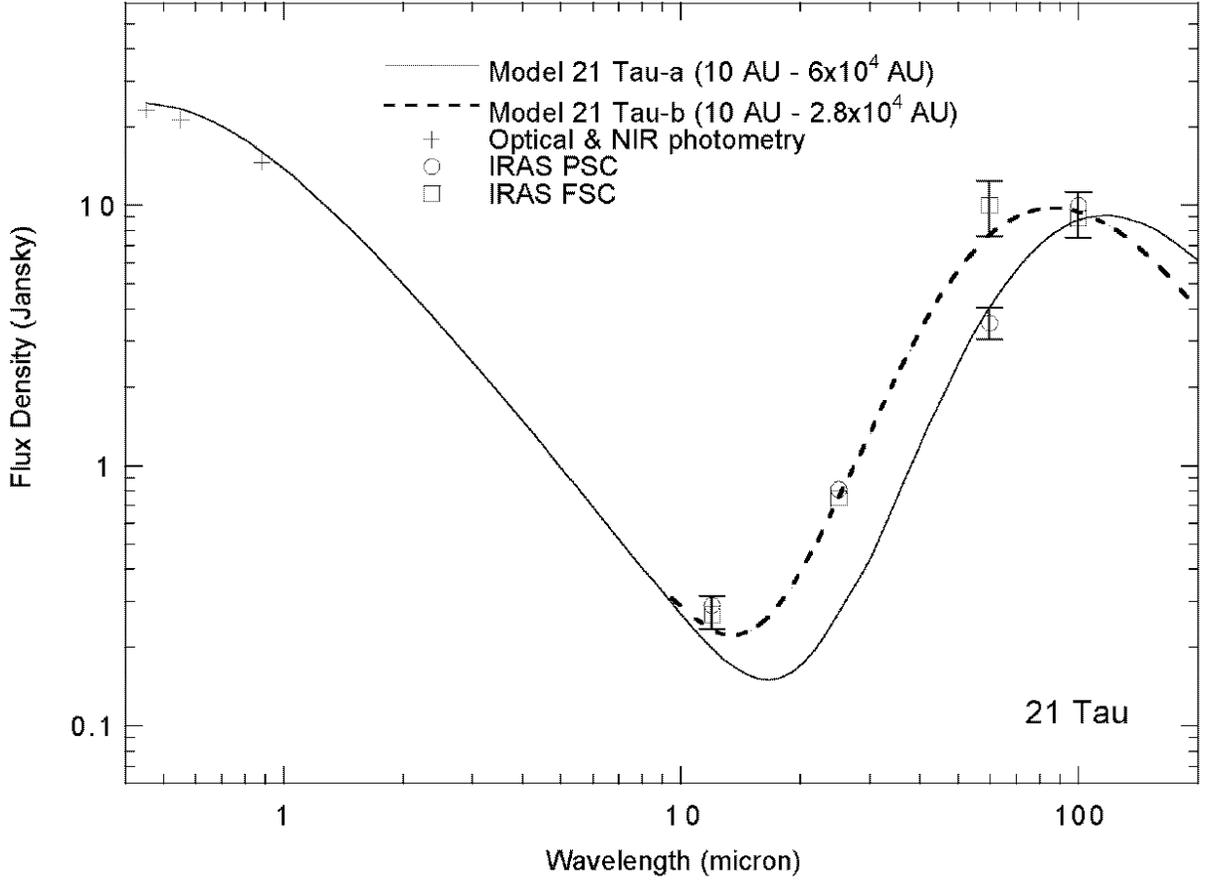}
\caption{Photometric data for 21 Tau (Table 4) and spectral energy distributions from
dust cloud models (Table 5).  Symbol sizes for the IRAS data approximate the sizes of error bars
unless marked otherwise.
The two models demonstrate different cloud properties that give fits to either the PSC or
FSC 100 $\mu$m data.  For the IRAS PSC data, the observed SED is similar to that of
18 Tau (Fig. 3).  Model 21 Tau-a has r$_{in}$=10 AU, but the hotter dust relative to
the 18 Tau models (Fig. 3) does not produce enough 12-25 $\mu$m flux to fit the observations.
A key difference in the IRAS FSC data for 21 Tau is the greater 60 $\mu$m flux density.  This permits
a fit to the entire SED with Model 21 Tau-b (Table 5).
\label{fig4}}
\end{figure}

\clearpage
\begin{figure}
\plotone{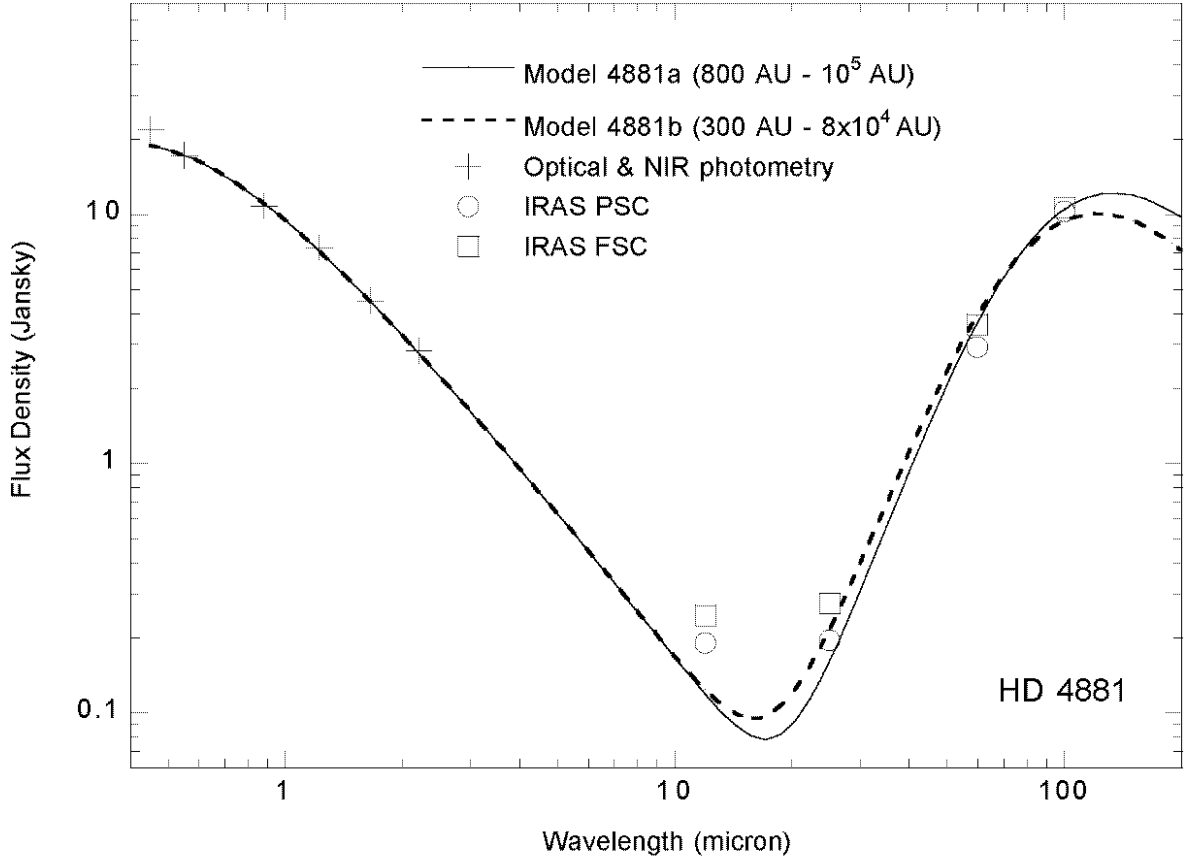}
\caption{Photometric data for HD 4881 (Table 4) and spectral energy distributions from
two dust cloud models (Table 5).  Symbol sizes for the IRAS data approximate the sizes of error bars. 
The two models demonstrate different cloud dimensions that give SED's 
consistent with the 25-100 $\mu$m data.
Model 4881-b demonstrates that decreasing outer radius relative to model 4881a diminishes the
100 $\mu$m emission.  Decreasing the inner radius in Model 4881-b enhances the 25 $\mu$m emission, but
with negligible effect on the 12 $\mu$m emission.
The 12 $\mu$m emission should be mostly photospheric, but the
IRAS data do not lie on the extrapolated photospheric blackbody curve.\label{fig5}}
\end{figure}

\clearpage
\begin{figure}
\plotone{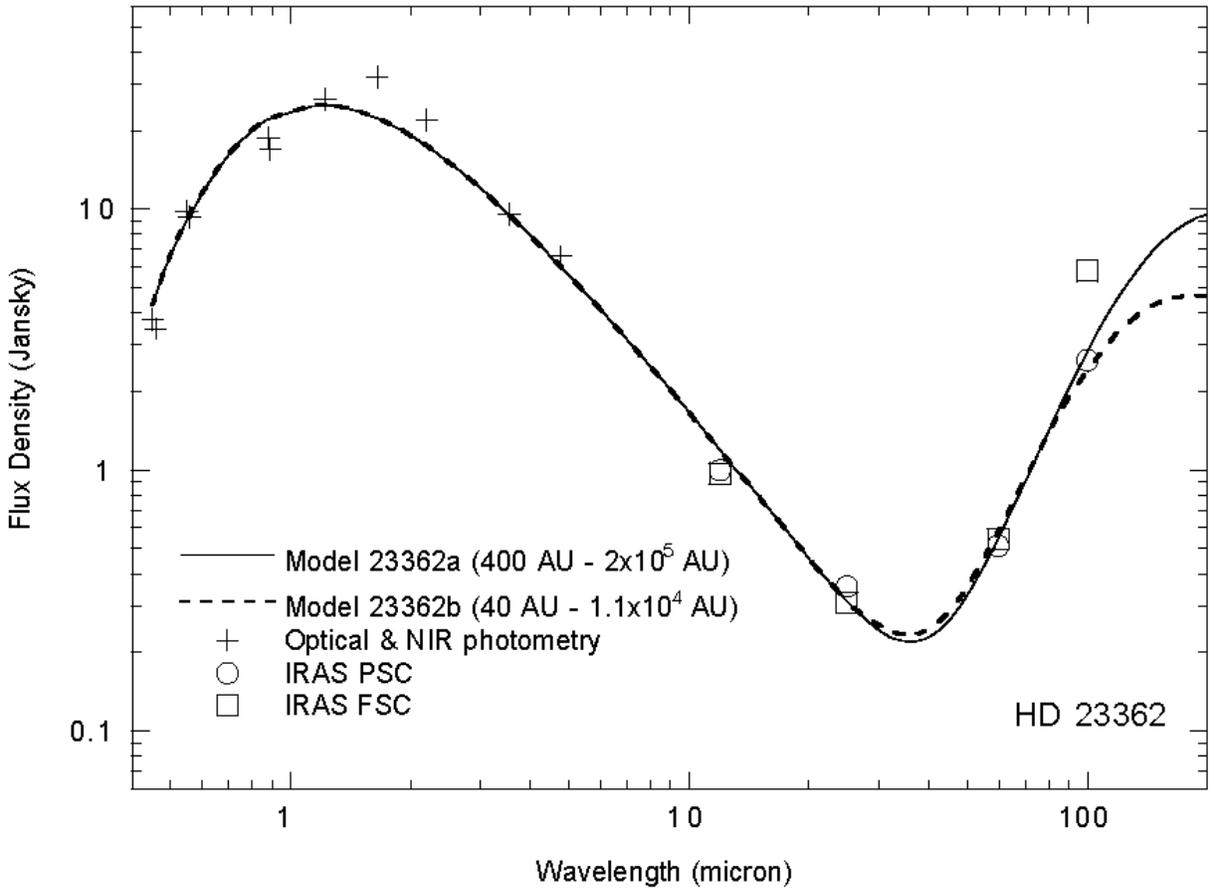}
\caption{Photometric data for HD 23362 (Table 4) and spectral energy distributions from two
dust cloud models (Table 5).  Symbol sizes for the IRAS data approximate the sizes of error bars. 
The two models demonstrate different cloud dimensions that fit the entire 12-100 $\mu$m PSC SED.
\label{fig6}}
\end{figure}

\clearpage
\begin{figure}
\plotone{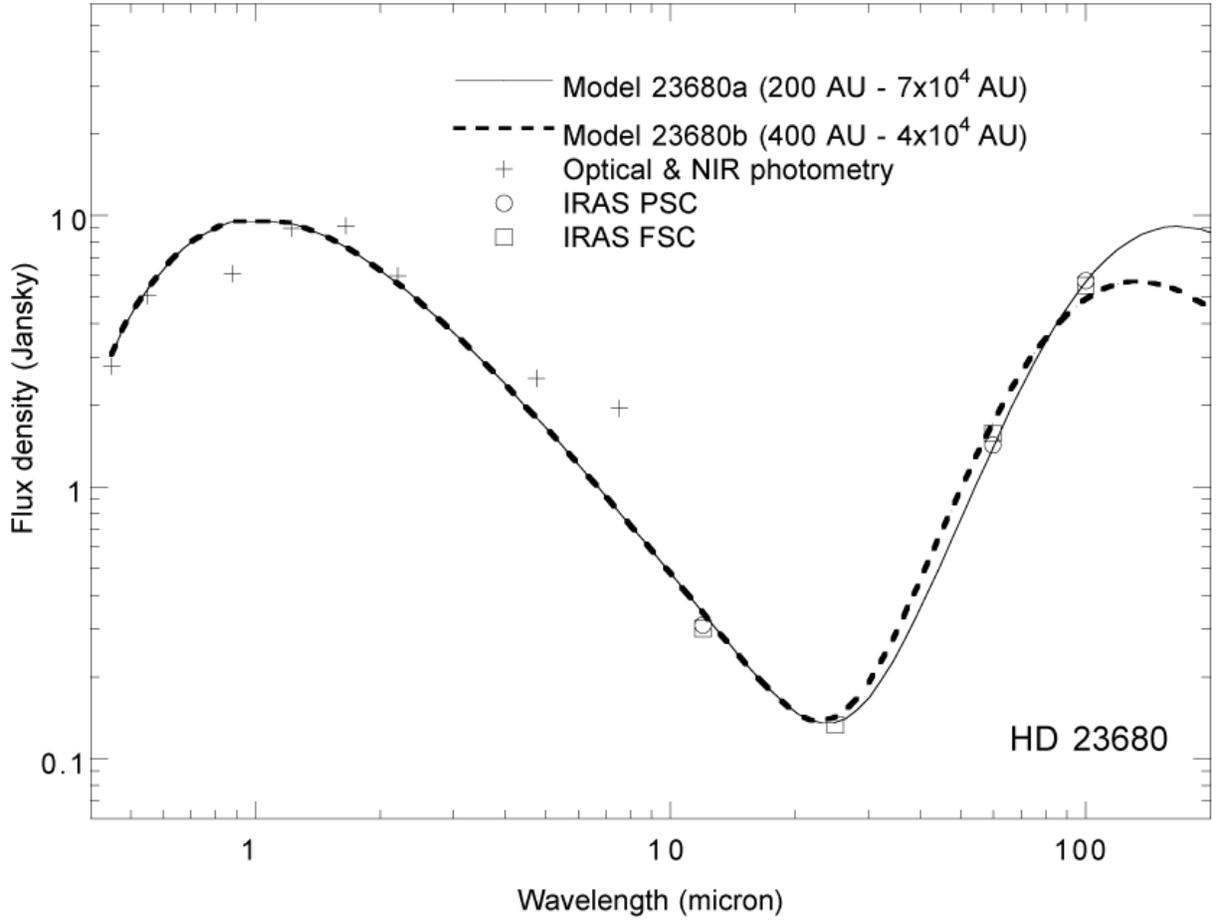}
\caption{Photometric data for HD 23680 (Table 4) and spectral energy distributions from
two dust cloud models (Table 5).  
Symbol sizes for the IRAS data approximate the sizes of error bars
unless marked otherwise.
The two models demonstrate different cloud dimensions that fit the entire 12-100 $\mu$m SED.
\label{fig7}}
\end{figure}

\clearpage
\begin{figure}
\plotone{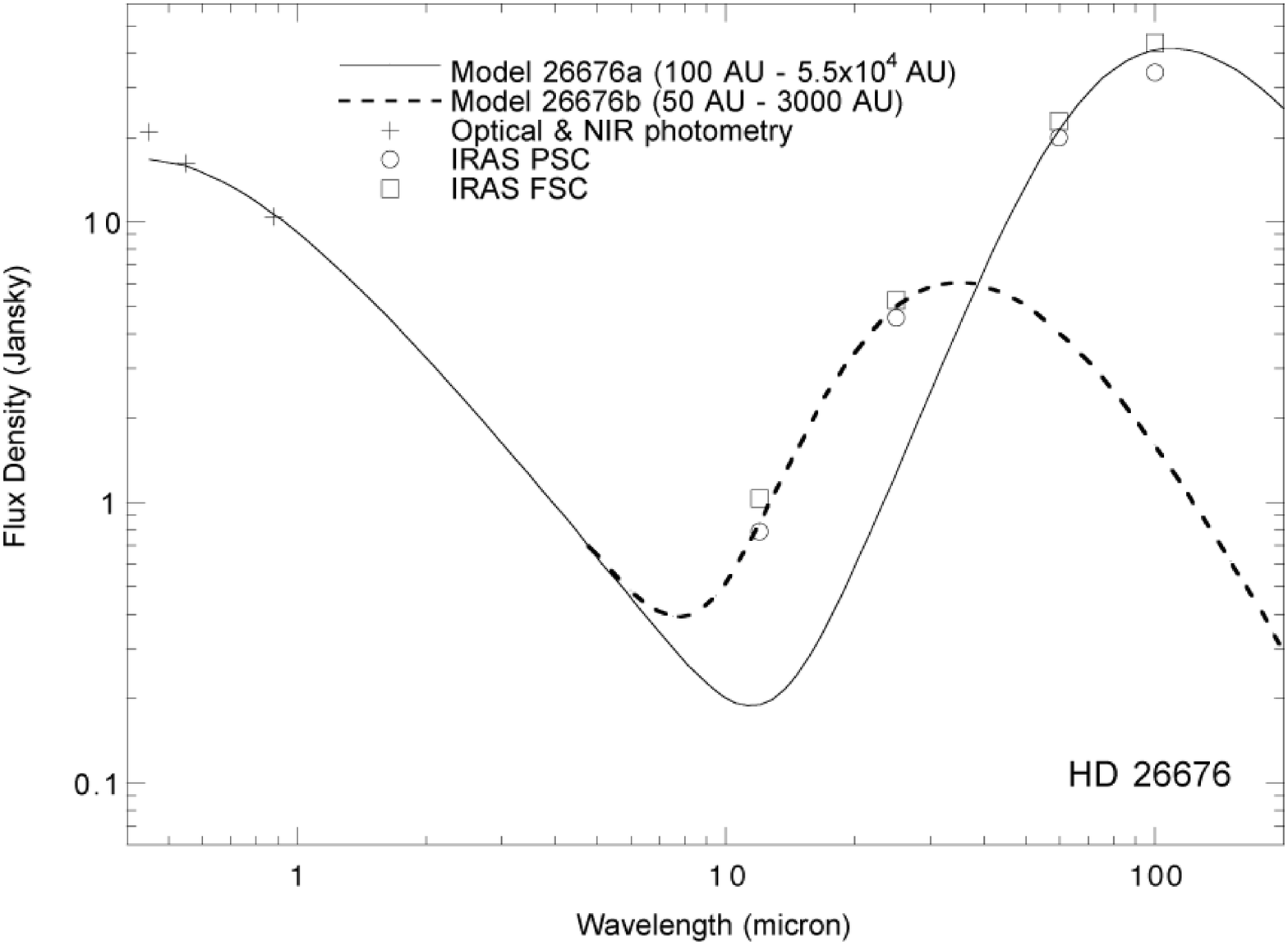}
\caption{Photometric data for HD 26676 (Table 4) and spectral energy distributions from
two dust cloud models (Table 5).  Symbol sizes for the IRAS data approximate the sizes of error bars. 
As with the Pleiads 18 Tau and 21 Tau (Figs. 3 and 4), no single model can simultaneously
fit the 12-25 $\mu$m and 60-100 $\mu$m regions of the SED simultaneously.
These two regions are fit independently by models 26676-a and 26676-b.
\label{fig8}}
\end{figure}

\clearpage
\begin{figure}
\plotone{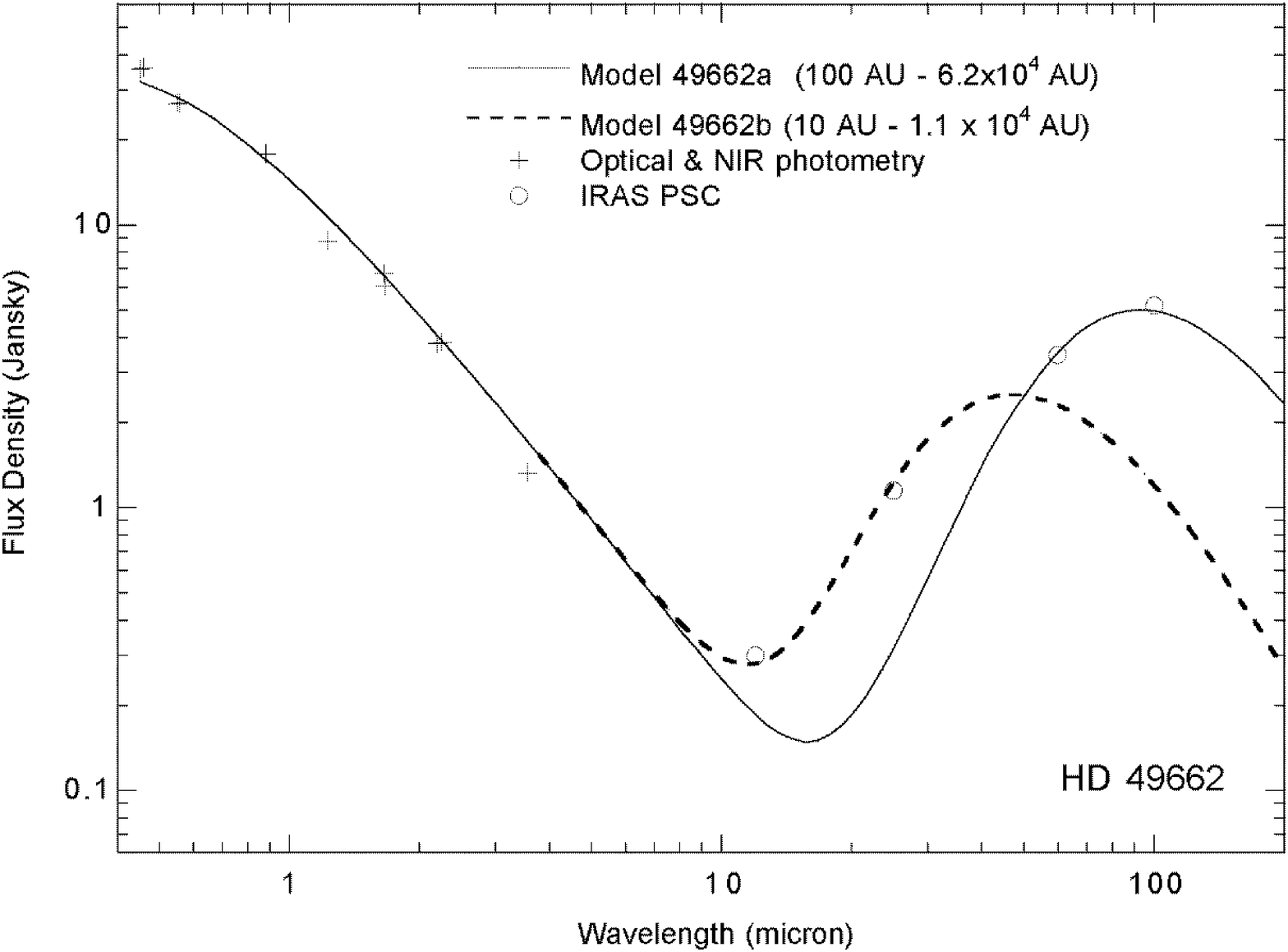}
\caption{Photometric data for HD 49662 (Table 4) and spectral energy distributions from
dust cloud models (Table 5). Symbol sizes for the IRAS data approximate the sizes of error bars. 
As with 18 Tau, 21 Tau, and HD 26676 (Figs. 3, 4 and 8), no single model can simultaneously
fit the 12-25 $\mu$m and 60-100 $\mu$m regions of the SED simultaneously.
These two regions are fit independently by models 49662-a and 49662-b (Table 5).
\label{fig9}}
\end{figure}

\clearpage
\begin{figure}
\plotone{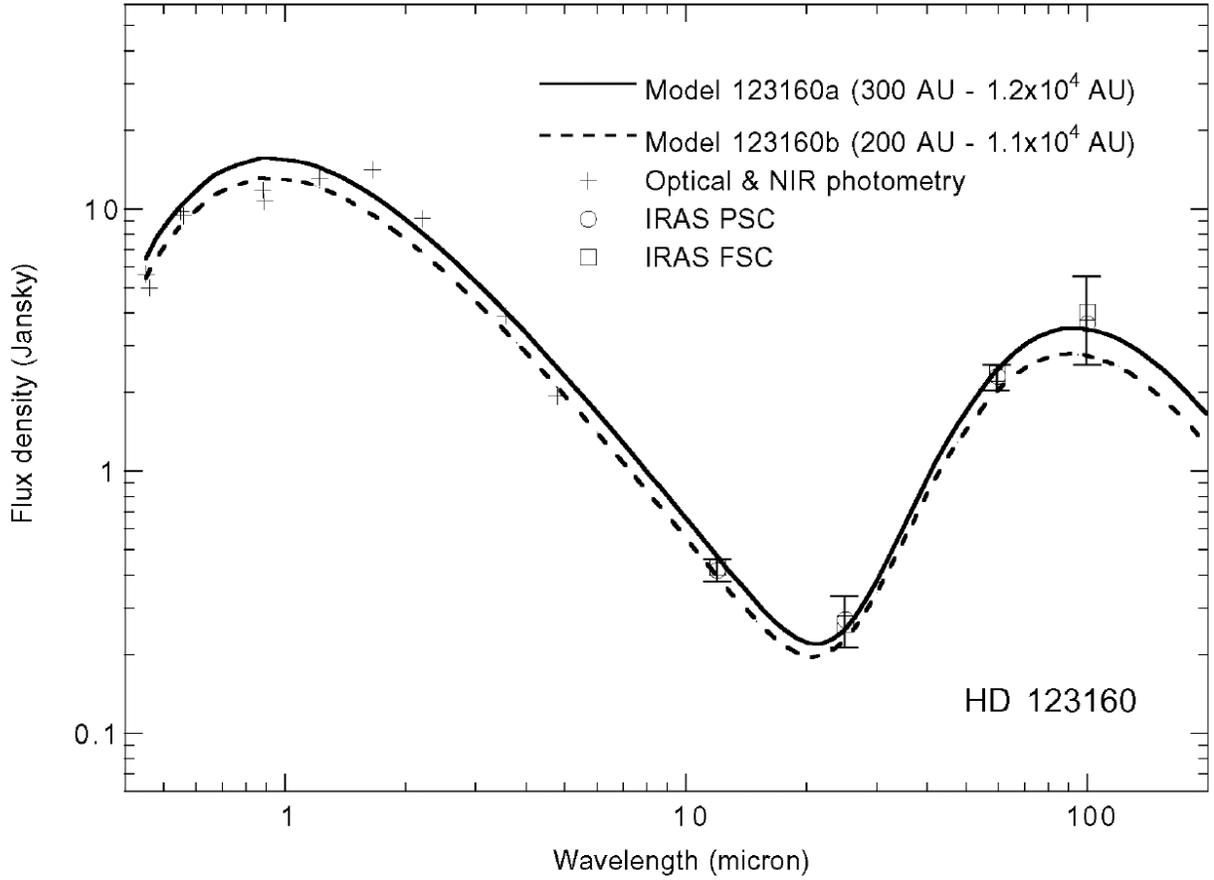}
\caption{Photometric data for HD 123160 (Table 4) and spectral energy distributions from
dust cloud models (Table 5).  Symbol sizes for the IRAS data approximate the sizes of error bars
unless marked otherwise.
The two models demonstrate different cloud radii that fit the data.  With model 123160-b, we
demonstrate that the distance could be increased to
d=120 pc and still obtain a satisfactory fit to the SED between 0.45 $\mu$m and 12 $\mu$m.
\label{fig10}}
\end{figure}

\clearpage
\begin{figure}
\plotone{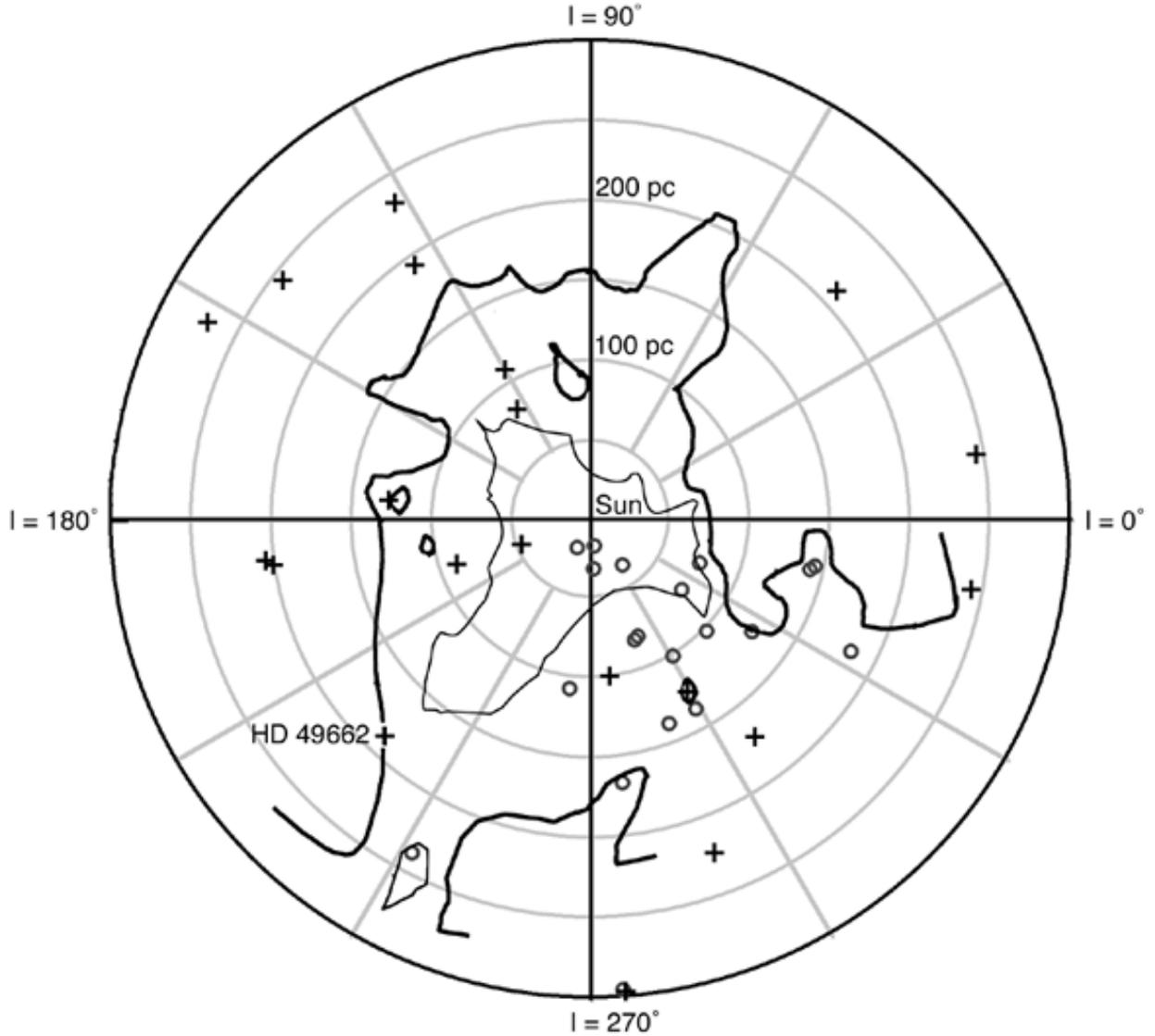}
\caption{Galactic plane (downward) view of Vega-like stars from 
Backman \& Paresce 1993 (crosses) and Mannings \& Barlow 1998 (circles).
The Sun is at the center.  Following \citet{sfe99},
only stars that have $-$18$\degr$ $<$ $b$ $<$ 18$\degr$ are plotted, and the Sun$-$star
distance is the $Hipparcos$ distance irrespective of the angle above or below the
reference plane.  The isocontours trace the 
NaI gas absorption from Fig. 3 in \citet{sfe99}.  The thin and thick contours correspond to
the 5 $m\AA$ and $>$50 $m\AA$ D2-line equivalent widths, respectively.
One exception is the closed contour near $l$=240$\degr$, d = 250 pc which
traces a 20 $m\AA$ equivalent width isocontour from \citet{sfe99}.  We show
it because the Vega-like source HD 52140 falls within the contour
boundary.  Two more examples of Vega-like stars associated with
a local overdensities of gas are HD 28149 at $l$=174$\degr$, d = 127 pc
and HD 108257 at $l$=299$\degr$, d = 123 pc.
\label{fig11}}
\end{figure}

\clearpage
\begin{figure}
\plotone{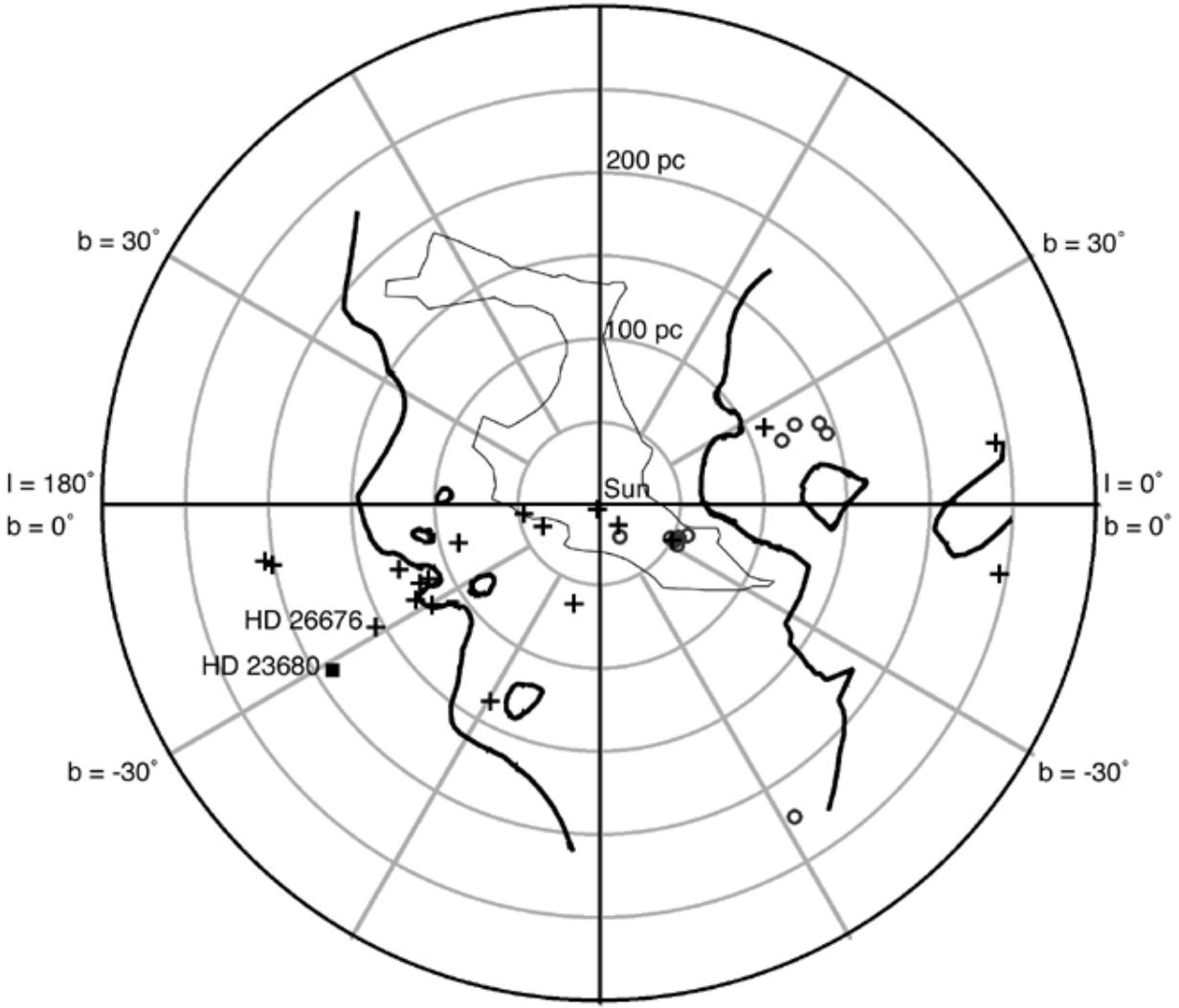}
\caption{Meridian plane view, perpendicular to the galactic
plane, and containing both galactic poles and the galactic
center to the right.  Symbols and contours same as Fig. 11.
Stars plotted if they fall within $l$=0$\degr\pm$18$\degr$ or $l$=180$\degr\pm$18$\degr$.
HD 23680 from our study falls within these limits and is plotted with a solid square.
\label{fig12}}
\end{figure}

\clearpage
\begin{figure}
\plotone{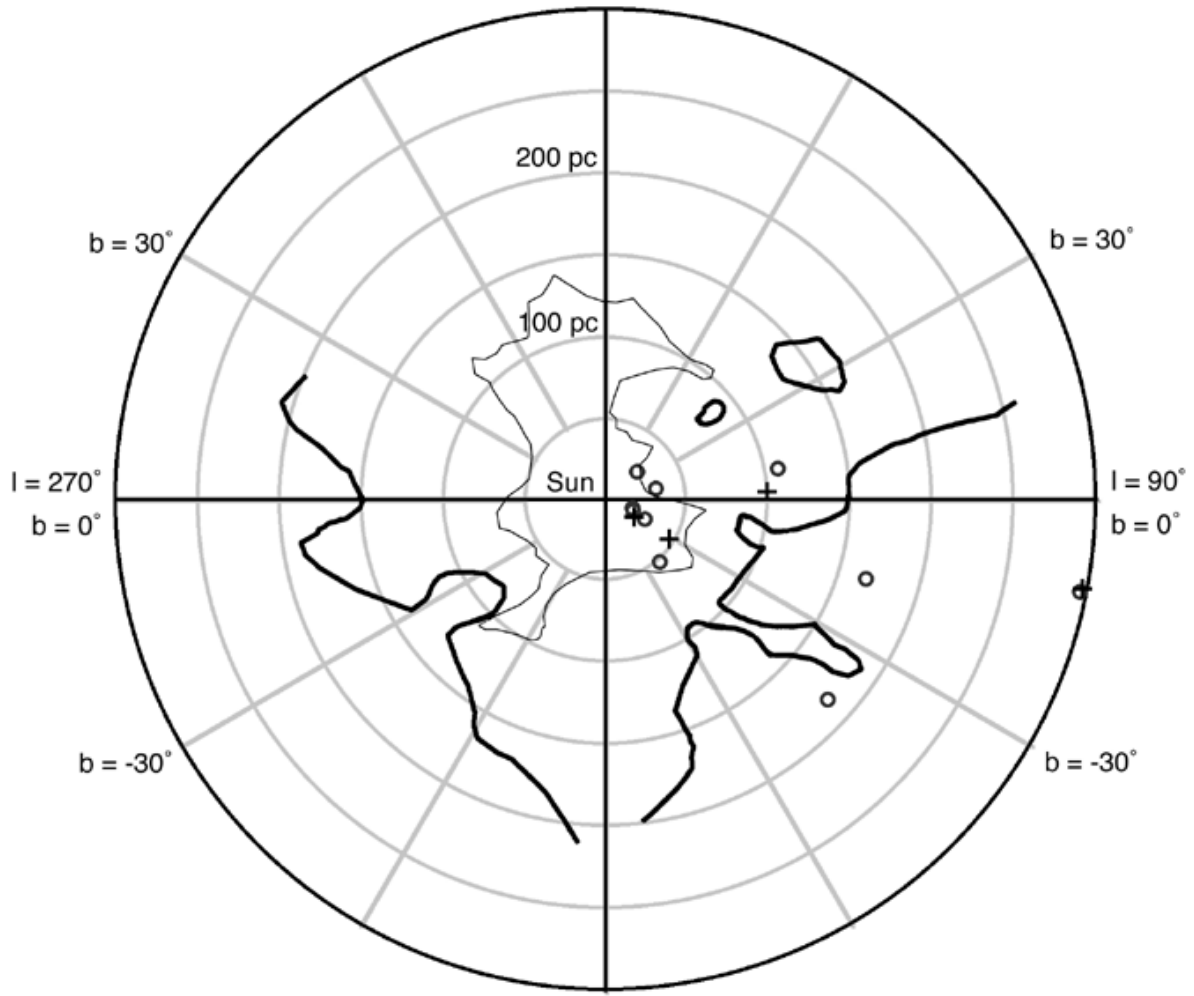}
\caption{Galactic rotation plane, perpendicular to both the galactic
plane and to the galactic center direction.  Symbols and contours same as Fig. 11.
Star plotted if they fall within $l$=90$\degr\pm$18$\degr$ or $l$=270$\degr\pm$18$\degr$.
\label{fig13}}
\end{figure}

\end{document}